\documentclass{aastex631}

\usepackage[utf8]{inputenc}
\usepackage[T1]{fontenc} 
\usepackage{comment}
\usepackage{xcolor}
\usepackage{epstopdf}
\usepackage{graphicx}
\usepackage{subfigure}
\DeclareUnicodeCharacter{2212}{-}

\begin{document}

\title{Resolved Dust Emission and CO Isotopologues in Giant Molecular Clouds of the Andromeda Galaxy}

\author{Chloe Bosomworth}
\affiliation{Centre for Astrophysics Research, University of Hertfordshire \\ College Lane, Hatfield \\ AL10 9AB, UK}
\affiliation{Center for Astrophysics, Harvard \& Smithsonian \\
60 Garden St, MS 72 \\
Cambridge, MA 02138, USA}

\author{Jan Forbrich}
\affiliation{Centre for Astrophysics Research, University of Hertfordshire \\ College Lane, Hatfield \\ AL10 9AB, UK}
\affiliation{Center for Astrophysics, Harvard \& Smithsonian \\
60 Garden St, MS 72 \\
Cambridge, MA 02138, USA}

\author{Charles J. Lada}
\affiliation{Center for Astrophysics, Harvard \& Smithsonian \\
60 Garden St, MS 72 \\
Cambridge, MA 02138, USA}

\author{Glen Petitpas}
\affiliation{Kavli Institute for Astrophysics and Space Research, Massachusetts Institute of Technology \\ Cambridge, MA 02139, USA}

\begin{abstract}

Dust emission at {submillimeter} wavelengths can be used to reliably trace the basic properties of molecular clouds. Early results {from} a recent Submillimeter Array (SMA) survey of the Andromeda Galaxy (M31) include the first detections of resolved dust continuum emission from individual giant molecular clouds (GMCs) in an external spiral galaxy. This paper updates {on the now-complete} SMA survey {of} 80 \textit{Herschel}-identified giant molecular associations (GMAs) in M31. The SMA survey simultaneously probes {dust continuum emission at} 230 GHz and the {$J = 2 \rightarrow 1$ transitions of the} CO isotopologues, $^{12}\rm CO$, $^{13}\rm CO$, and $\rm C^{18}O$ at a spatial resolution of $\lesssim 15~\mathrm{pc}$. Dust continuum emission was detected in 71 cloud cores, of which {26 were} resolved. This more than doubles the size of the previous sample. By comparing dust and CO observations {with} identical astrometry, we directly measure the dust mass-to-light ratios, $\rm \alpha^{\prime}_{^{12}CO}$, {and} $\rm \alpha^{\prime}_{^{13}CO}$. We derive $<\alpha^{\prime}_{\rm ^{12}\rm CO}>~=~0.070~\pm~0.031~M_{\odot}\,(\rm K~km~s^{-1}~pc^{2})^{-1}$ and $<\alpha^{\prime}_{\rm ^{13}\rm CO}>~=~0.37~\pm~0.20~M_{\odot}\,(\rm K~km~s^{-1}~pc^{2})^{-1}$ for the increased sample, {which are} in agreement with previously reported values. From virial analysis, we find that 80\% of the GMC regions traced by resolved dust emission are bound and {close to virial equilibrium}. Finally, we update our analysis {on} the metallicity dependence of $\rm \alpha^{\prime}_{\rm CO}$ by combining SMA observations with existing MMT/Hectospec optical spectroscopy {toward} H~{\sc ii} regions. We {find} no trend {in} $\rm \alpha^{\prime}_{\rm CO}$ with metallicity, supporting the previous findings.

\end{abstract}


\section{Introduction} \label{sec:intro}

Giant molecular clouds (GMCs) are the main sites of star formation. {The} physical conditions within GMCs {determine} the star formation rate (SFR), the primary metric for studies of galaxy (e.g., \citealp{Saintonge_2022}) and cosmic evolution (e.g., \citealp{Madau_2014, Tacconi_2020}). Molecular gas provides the fuel for star formation, and the most abundant molecule in the cold interstellar medium (ISM), which is largely contained within GMCs, is molecular hydrogen ($\rm H_{2}$), making up 75\% {of the mass}. Additionally, GMCs are composed of {approximately} 24\% atomic helium (He) and 1\% interstellar dust by mass \citep{Saintonge_2022}. Both $\rm H_{2}$ and He are difficult to detect at the cold temperatures typical of GMCs{; therefore, we rely on} rarer molecules to detect and probe the molecular gas. In particular, CO is primarily used in extragalactic studies (see, e.g., \citealp{Bolatto_2013} for a review) because it emits the strongest molecular emission lines in cold GMCs. However, CO only accounts for $\sim0.03\%$ of GMC mass (e.g., \citealp{Heyer_2015}), is optically thick {in dense cloud regions} due to its high opacity, and depletes {onto} dust grains. As a result, CO traces the diffuse GMC envelopes and {does not} reflect the denser regions of the cloud. Dense gas is key to understanding global star formation efficiency (SFE) and the star formation rate (SFR; \citealp{Gao_2004, Lada_2010, Lada_2012}), and so it is vital to investigate alternative tracers that better probe {the physical properties of clouds}.

Observations of dust extinction in Milky Way (MW) GMCs have provided detailed information on cloud structure {as well as} the first robust measurements of GMC masses (e.g., \citealp{Lada_1994, Alves_2001, Goodman_2009, Lada_2010}). Dust is $\rm \sim30 \times$ more abundant than CO in GMCs, and since molecules such as $\rm H_{2}$ and CO form on the surface of dust grains (e.g., \citealp{Wakelam_2017}), it is well mixed with molecular gas. At {submillimeter} wavelengths, we can probe thermal dust continuum emission, which traces $\rm H_{2}$ gas properties more accurately than CO due to its lower optical depth (e.g., \citealp{Scoville_2016, Scoville_2017}). Although dust extinction measurements are independent of dust temperature ($T_{\rm dust}$), dust emission depends on $T_{\rm dust}$ and opacity ($\kappa_{\nu}$). However, unlike molecular excitation lines, dust emission does not {depend on} excitation conditions.

To fully understand the star formation processes and conditions {in} a galaxy, we need to probe GMCs across the entire {disk of the galaxy}. This is difficult {in the MW because of distance measurement uncertainties and line-of-sight dust extinction, particularly for clouds located near the center of the galaxy.} GMCs within an external galaxy are approximately the same distance from us, and therefore we can obtain a large sample of measurements at {a} consistent resolution. As the most easily observable tracer, CO is primarily used to trace molecular gas in extragalactic GMCs. The first resolved CO detections of extragalactic GMCs were {made in} the Andromeda Galaxy \citep{Vogel_1987, Lada_1988}. More recently, resolved observations of extragalactic GMCs in CO on {$\sim$100 pc scales} (e.g., PHANGS-ALMA; \citealp{Leroy_2021}) and {$\sim$10 pc scales} (e.g., \citealp{Colombo_2014, Faesi_2018}) {have become routine}.

\textit{Herschel} studies reveal the spatial distribution of giant molecular associations (GMAs) in M31, traced by dust emission at {$\sim$93~pc} spatial resolution \citep{Fritz_2012, Kirk_2015}. These GMAs have sizes of {$\sim$80--300~pc}, and thus {typically} encompass associations of {multiple} unresolved GMCs. Generally, GMCs have sizes of {$\sim 20$--100~pc} \citep{Solomon_1979}, which means that individual GMCs are not resolved by \textit{Herschel}. Extragalactic studies of resolved GMCs are predominantly limited to CO due to the high sensitivity required to detect and resolve dust emission from extragalactic GMCs. This changed with the recent wideband sensitivity upgrade of the Submillimeter Array (SMA; \citealp{Grimes_2016}), which has enabled the first resolved dust emission detections from GMCs in an external spiral galaxy, M31, at {$\rm \lesssim 15~pc$} \citep{Forbrich_2020, Viaene_2021}.

To detect resolved dust continuum from individual GMCs in M31, long exposure times were required. Therefore, observing the entire galaxy would be extremely difficult, and {as a result, a sample of} targets {was} selected from the HELGA dust continuum survey of {\it Herschel}-identified GMAs \citep{Kirk_2015}. The aforementioned SMA survey obtained simultaneous dust emission and three CO isotopologues: $\rm ^{12}CO(2-1)$, $\rm ^{13}CO(2-1)$, and $\rm C^{18}O(2-1)$, with identical $(u, v)$ coverage, spatial scale{s}, and calibration (see \citealp{Forbrich_2020, Viaene_2021, Lada_2024}). Since M31 GMCs have an approximately uniform distance from us ({$\sim780~\rm kpc$}: \citealp{Stanek_1998}), this dust survey has the advantage of eliminating significant uncertainties in distance measurements present in MW studies.

Two previous papers report 32 dust continuum detections (10 resolved) associated with M31 GMCs from the analysis of the first two observing runs of an SMA large survey \citep{Forbrich_2020, Viaene_2021}. In this paper, we present the updated analysis of the now completed survey after four observing runs, increasing the sample size to 71 dust continuum detections (26 resolved). The sensitivity of these observations enabled resolved dust emission detections on scales of {$\sim 15$~pc}. Since the sensitivity to dust in these observations is not sufficient to trace entire GMCs, individual dust detections are hereafter referred to as ``dust cores'' (following \citealp{Viaene_2021}). {The sensitivity achieved in $^{12}$CO enables measurement of the full $^{12}$CO(2-1) emitting area of each GMC \citep{Viaene_2021}, allowing us to identify the dust emission associated with these clouds.}

From CO and dust emission {in} the same clouds, we can directly calculate $\alpha^{\prime}_{\rm CO}$, the ratio between dust mass ($M_{\rm dust}$) and CO luminosity ($L_{\rm CO}$). This {ratio is related to} the CO-to-$\rm H_{2}$ conversion factor, $\alpha_{\rm CO}$, {through} the gas-to-dust ratio ($R_{\rm g/d}$). Therefore, calculating $\alpha_{\rm CO}$ requires an assumption of $R_{\rm g/d}$, for which the typical MW value is $R_{\rm g/d} \sim 136$ (including a factor of \textcolor{black}{1.36} to account for He; \citealp{Draine_2007}). {However,} chemical evolution models (e.g., \citealp{Mattsson_2012, Hirashita_2017}) and observations (e.g., \citealp{Sandstrom_2013, Giannetti_2017}) have shown that $R_{\rm g/d}$ can vary with dust grain size and metallicity. By measuring $\alpha^{\prime}_{\rm ^{12}CO}$ and $\alpha^{\prime}_{\rm ^{13}CO}$ independently of $R_{\rm g/d}$, we exploit the full capability of dust as a molecular gas tracer and gain insight into the ISM conditions of star formation directly from these observations. {Moreover, by deriving $\alpha^{\prime}_{\rm CO}$ for both the $\rm ^{12}CO$ and the rarer CO isotopologue $\rm ^{13}CO$ in the same sources, we can compare molecular gas components with different opacities.}

The combination of CO and dust continuum emission observations with identical astrometry enables the calculation of both the luminous and virial masses of these dust cores. \citet{Lada_2024} evaluate the Larson paradigm, which states that GMCs are in dynamical equilibrium with gravity, for a sample of 117 GMCs in M31. Using data from the same SMA survey described in this work and employing $\rm ^{12}CO(2-1)$ to trace cloud boundaries, \citeauthor{Lada_2024} {found} that this paradigm does not hold for {most M31 GMCs}. {In contrast,} for the dense regions of GMCs traced by $\rm ^{13}CO$ emission in these observations, the majority were found to be gravitationally selfbound. These $\rm ^{13}CO$ detections are referred to as `clumps' because they, like dust, do not trace entire GMCs, and a single GMC can have multiple associated clumps. In this paper, we assess the virial state of resolved dust cores, which are constrained to the dense regions of GMCs in our observations.

The accuracy of CO as a molecular gas tracer depends on correctly accounting for the effects of metallicity on CO abundance and conversion factors. CO conversion factors, $X_{\rm CO}$ and $\alpha_{\rm CO}$, are predicted to vary with metallicity (e.g., \citealp{Leroy_2011, Bolatto_2013}). At lower metallicities, less carbon (C) and oxygen (O) {are} available, resulting in lower CO abundance \citep{Maloney_1988, Israel_1997}. Therefore, $L_{\rm CO}$ per unit gas mass is predicted to decrease at lower metallicities. Furthermore, {approximately} 30--50\% of metals {are expected to} deplete onto dust grains \citep{Savage_1996, Draine_2007}, so $M_{\rm dust}$ and dust opacity also increase with metallicity. Dust extinction shields $\rm H_{2}$ and CO molecules from photodissociation {by} UV radiation.

Therefore, more effective dust shielding at higher metallicities allows CO to trace a larger fraction of the molecular gas (e.g., \citealp{Glover_2010}, \citeyear{Glover_2012}). The metallicity dependence of $\alpha^{\prime}_{\rm CO}$ provides insight into the relative effects of metallicity on $M_{\rm dust}$ and $L_{\rm CO}$, and how this {in turn} could affect $\alpha_{\rm CO}$ if $R_{\rm g/d}$ is assumed to be constant. In \citet{Bosomworth_2025}, an approximately constant $\alpha^{\prime}_{\rm ^{12}CO}$ was found {across} the metallicity range 12~+~log(O/H)~=~8.4--8.7. A possible explanation {for this behavior} is that the simultaneous effects on $M_{\rm dust}$ and $L_{\rm CO}$ may cancel out at these metallicities. By comparing our $\alpha^{\prime}_{\rm ^{12}CO}$ values from the updated dust core sample with gas-phase oxygen abundance measurements (O/H) of individual H~{\sc ii} regions associated with a subset of these GMCs, we update the result presented in \citet{Bosomworth_2025}. This {allows} us to further constrain the metallicity dependence of $\alpha^{\prime}_{\rm CO}$.

This paper is structured as follows. In Section~\ref{sec:obs}, we describe our SMA observations and {the selection of our target sources}. In Section~\ref{sec: Methodology}, we outline our methodology and data analysis procedures. In Section~\ref{sec:Res}, we present the results of our analysis, including $L_{\rm CO(2-1)}$ for both $\rm ^{12}CO$ and $\rm ^{13}CO$, along with the corresponding values of $M_{\rm dust}$ and $\alpha^{\prime}_{\rm CO}$. {Additionally, we perform} a virial analysis of {the} dust cores and {investigate} the metallicity dependence of $\alpha^{\prime}_{\rm ^{12}CO}$. Finally, in Section~\ref{sec: sum}, we present our summary and conclusions.

\section{Observations} \label{sec:obs}

Observations were obtained from an SMA large program targeting individual resolved GMCs in M31. This survey {used} the SMA in the subcompact configuration, which achieved a synthesized beam size of 4.\!$\farcs$5~$\times$~3.\!$\farcs$8 ($\sim$15~pc). Occasionally, one or two antennas were out of service due to technical issues, {which increased the typical beam size to $\sim$8\arcsec~$\times$~5\arcsec~($\sim$30~$\times$~19~pc), still sufficient to resolve individual GMCs.} At a frequency of 230~GHz (1.3~mm), the bandwidth was 32~GHz for the first two observing runs and was subsequently {increased} to 48~GHz for the final two runs. Within this frequency range, we have access to thermal dust continuum emission and the $^{12}\rm CO(2\text{--}1)$, $^{13}\rm CO(2\text{--}1)$, and $\rm C^{18}O(2\text{--}1)$ molecular lines. Analysis and results from the first two observing runs are presented by \citet{Forbrich_2020} and \citet{Viaene_2021}.

\begin{figure}
    \centering
    \includegraphics[width=0.95\linewidth]{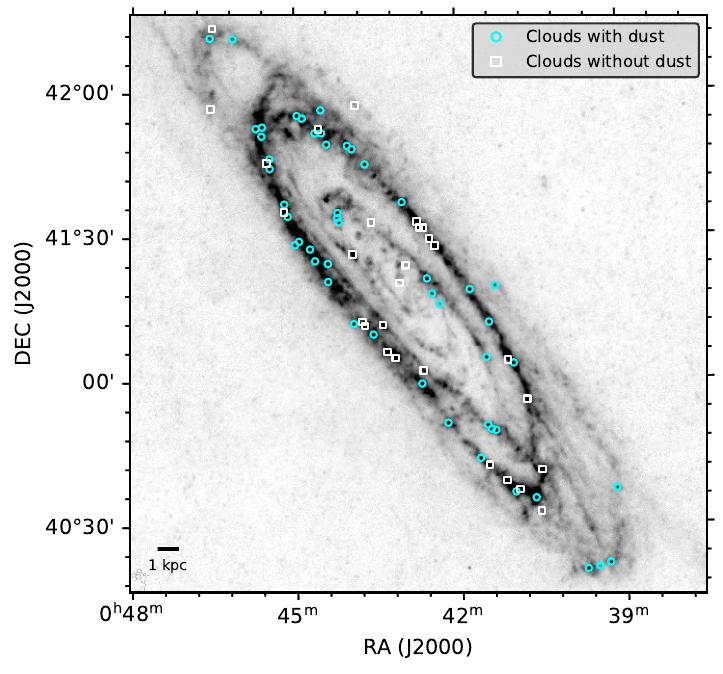}
    \caption{{Locations of SMA-observed GMAs \citep{Kirk_2015} over-plotted onto the \textit{Herschel} Spectral and Photometric Imaging Receiver (SPIRE) 500-micron image \citep{Fritz_2012}, with GMAs for which we obtained dust detections displayed as open circles, and GMAs observed but with no dust detections displays as filled circles. The 1 kpc scalebar assumes the distance to M31 to be 780 kpc; \citep{Stanek_1998}.}}
    \label{fig:Obs}
\end{figure}

As the entire galaxy could not be observed within a reasonable timeframe, targets were selected from the HELGA dust continuum survey of {\it Herschel}-identified GMAs \citep{Kirk_2015}. The targets were chosen to cover a representative range of {\it Herschel} fluxes and physical cloud properties present throughout the M31 {disk}. {Toward the beginning of the survey, it was found} that the likelihood of detecting dust continuum emission on $\sim$15~pc scales was not strongly correlated with {\it Herschel} flux on $\sim$100~pc scales, {nor was it dependent on position within the disk \citep{Forbrich_2020}}. {The spatial distribution of the SMA-observed GMAs across M31 are displayed in Figure \ref{fig:Obs}, overlaid on the Spitzer MIPS Infrared 24 microns image \citep{Gordon_2006}. These observations produced dust detections from GMCs distributed across the disk of M31, spanning a wide range of galactocentric radii and {\it Herschel} fluxes}.

Data were collected from 100 pointings toward 80 targets over four fall seasons from 2019 to 2022. Each source was observed for $\sim$6 hours{,} and the target sensitivity of 0.20~mJy~beam$^{-1}$ {to dust continuum} was consistently {achieved}. The typical rms values reached for the $\rm ^{12}CO(2-1)$ and $\rm ^{13}CO(2-1)$ molecular lines respectively are 20~mJy~beam$^{-1}$ and 14~mJy~beam$^{-1}$ (0.04~K and 0.03~K), per $\sim\Delta1\rm~km~s^{-1}$ channel. {We obtained} dust continuum detections associated with GMCs for 51 of the targeted GMAs. A typical resolution of $\sim$15~pc was achieved. Two receivers were used{,} RxA and RxB, tuned to local oscillator frequencies of 225.55~GHz and 233.55~GHz{,} respectively. This provided a CO resolution of 140.0~kHz per channel, and initially a continuous bandwidth between 213.55 and 245.55~GHz. Within this frequency range, the emission lines $\rm ^{12}CO(2-1)$, $\rm ^{13}CO(2-1)$, and $\rm C^{18}O(2-1)$ are accessible. Subsequently, the coverage was extended to 48~GHz, providing continuous coverage between 209.5~GHz and 249.5~GHz, with an overlap of 8~GHz.

These data were obtained in the form of spectral cubes, from which the CO-contaminated channels were extracted and analyzed using the Millimeter Interferometer Reduction Interactive Data Language (MIR~IDL) software for the SMA. Continuum images were computed from the average of the remaining channels. For $\rm ^{12}CO(2-1)$ and $\rm ^{13}CO(2-1)$, integrated-intensity (moment~0) maps were produced, as well as data cubes containing the corresponding velocity information. Raw spectra were binned to a velocity resolution of either 1 or 1.5~km~s$^{-1}$. Additional spectral flagging, {inversion}, and CLEANing were performed in Miriad\footnote{cfa.harvard.edu/sma/miriad/}. For detailed data calibration and reduction {procedures}, we refer {the reader to} \citet{Forbrich_2020}.

\section{Methodology} \label{sec: Methodology}

In this section, we outline the extraction of thermal dust emission detections from the continuum images and refine the final sample used in further analysis.  
The sample presented here includes those previously reported in \citet{Viaene_2021}, {along with} new observations and updated measurements obtained using the methodology outlined below. We make use of both dust continuum images and CO moment~0 maps to {ensure that our detected dust continuum cores are associated with GMCs}. In addition, CO velocity information is used to extract spectral line profiles for $L_{\rm CO}$ measurements. For the analysis of GMCs and $^{13}\rm CO$ clumps, as defined by their CO emission contours from this same dataset, we refer to \citet{Lada_2024}.

\subsection{Dust Core Extraction}

To identify and extract individual GMCs and dust cores from each observation, we define contours based on the image rms, $\sigma$. The high sensitivity achieved in the $^{12}\rm CO(2-1)$ data means that the 3$\sigma$ contours {typically} trace the entire CO-emitting area of each GMC \citep{Viaene_2021}. Because {the} dust continuum images and CO moment~0 maps have identical astrometry, we no longer depend on the global image statistics to identify dust continuum emission. Instead, {we} perform a targeted search within {the} GMC boundaries. We employ a 2.5$\sigma$ detection threshold, lower than the {3$\sigma$ threshold used} by \citet{Forbrich_2020} and \citet{Viaene_2021}. A 2.5$\sigma$ threshold improves our sensitivity to dust emission compared to previous analyses and introduces only minimal contamination from unrelated foreground or background sources. Applying this criterion to the entire survey, including the observations previously analyzed by \citet{Forbrich_2020} and \citet{Viaene_2021}, {more than doubles} the size of the previous sample.

At {the achieved} sensitivity, dust emission and $^{13}\rm CO$ do not trace the entire {extent of each GMC, and} we find that {a single} GMC can have multiple associated dust cores and $^{13}\rm CO$ clumps. Dust continuum emission at the 2.5$\sigma$ level was detected in 35 targeted GMAs, associated with 57 {separate} GMCs. Compared to the previous sample from \citet{Viaene_2021}, the complete sample of dust cores has increased from 32 to 71, and the number of resolved cores has increased from 10 to 26.

We define a resolved core as one which has a 2.5$\sigma$ contour at least 20\% larger than the synthesized beam, and {analyze} these separately from the unresolved sources. This allows us to evaluate the extent to which measurements may be affected by differences in the beam-filling factors of CO and dust. Finally, pixels outside the half-power primary beam ($r = 27\farcs5$, or $\sim$100~pc) are removed, except for one source (K098Bd1; see Figure~\ref{fig: Atlas}) where we find a resolved dust detection associated with a GMC {whose dust and CO contours both} partially overlap the edge of the half-power primary beam.

\subsection{Dust Mass Measurements} \label{sec:DustMass}

To calculate the dust mass of an individual core, $M_{\rm dust}$, we use the modified blackbody emission model of \citet{Hildebrand_1983}:

\begin{equation}
    M_{\rm dust} = \frac{S_{\nu} D^{2}}{\kappa_{\nu} B_{\nu}(T_{\rm dust})}.
\end{equation}

Here, $D = 780$~kpc \citep{Stanek_1998}, and $S_{\nu}$ is the continuum flux {density} measured within the 2.5$\sigma$ contour. We {adopt} a dust opacity of $\kappa_{\nu} = 0.0425~\rm m^{2}\,kg^{-1}$ from the THEMIS dust model \citep{Jones_2017} at 230~GHz, which has been calibrated using {\it Planck} data and the MW dust constraints of \citet{Ysard_2015}. This dust model has previously been shown to perform well in radiative transfer modeling of M31 by \citet{Viaene_2017}. Finally, we assume a dust temperature of $T_{\rm dust} = 18$~K, the median value derived for the \citet{Kirk_2015} GMAs.

To measure $M_{\rm dust}$ from our continuum images, we must first calculate $S_{\nu}$ for each dust core. For resolved sources, we calculate the average flux from all pixels within the {2.5$\sigma$ contour and multiply this value by the source area (in pixels)} to obtain $S_{\nu}$. For unresolved sources, we use the peak pixel flux and multiply it by the beam area (in pixels).

There is a possible uncertainty associated with the assumption of a constant dust temperature; therefore, the magnitude of this effect was previously assessed by \citet{Viaene_2021}. By assuming constant dust temperatures of $T_{\rm dust} = 15$~K and $T_{\rm dust} = 25$~K (the range of dust temperatures found for {GMAs studied by} \citealp{Kirk_2015}) in the dust mass calculations, \citet{Viaene_2021} found that a {$\pm5~\rm K$} variation in dust temperature does not significantly affect the results.

\subsection{CO Luminosity measurements}
\label{Sec: COlum}

{To calculate the CO luminous mass, $M_{\rm lum}$, we use CO line profiles extracted from the same regions of each GMC as the dust emission.} We extract line profiles for $\rm ^{12}CO(2-1)$ and $\rm ^{13}CO(2-1)$ over a {$100~\rm km~s^{-1}$-wide} band, centered on the {line peak}, and computed within the boundaries of each individual dust core. In the case of resolved dust cores, we extract the mean line profile of all pixels within the 2.5$\sigma$ dust core boundary. For unresolved dust cores, we {instead use} the line profile of the peak pixel within the contour. Line profiles are well resolved in velocity, and their line widths are comparable to {those of MW} GMCs (e.g., $\sim4$–$6~\rm km~s^{-1}$; \citealp{Rice_2016}), except for source K008, which contains two dust cores with extracted $\rm ^{12}CO$ line widths of $\sim18$ and $\sim31~\rm km~s^{-1}$. The CO intensity, $I_{\rm CO}$, of each dust core is measured from a Gaussian fit to the line profile{. We then calculate} $L_{\rm CO}$ following {Equation \ref{eqn: LCO}},

\begin{equation}
\label{eqn: LCO}
    L_{\rm CO} = I_{\rm CO} \times A_{\rm source}, 
\end{equation}

{where $A_{\rm source}$ is in pixels, and corresponds to the 2.5$\sigma$ contour area for resolved sources, and the beam area for unresolved sources.}

Previous $L_{\rm CO}$ measurements reported by \citet{Forbrich_2020} and \citet{Viaene_2021} were derived from moment~0 maps. Unlike the dust emission, our $^{12}\rm CO(2-1)$ observations are not limited by sensitivity, and the emission is detected in all moment~0 images. Given the strength of the $^{12}\rm CO(2-1)$ emission, regions of images that are free of $^{12}\rm CO$ emission are much less common than for dust and $^{13}\rm CO$ emission. This makes an accurate determination of the $^{12}\rm CO$ image rms difficult. The updated method used in this work improves the uncertainty estimate for $I_{\rm CO}$, which is calculated following:

\begin{equation}
    \delta I_{\rm CO} = \sigma_{\rm channel}\,\Delta v\, \sqrt{N_{channel}}
\end{equation}

where $\sigma_{\rm channel}$ is the rms noise per channel calculated from two $\rm 20~km\,s^{-1}$ ranges on both sides of the CO line profile, $\Delta v$ is the channel width (1 or 1.5~km\,s$^{-1}$) and $N$ is the number of channels summed over the line profile (FWHM~$\times 2.4$). This ratio is multiplied by the dust core area (resolved sources) or the beam area (unresolved sources) to calculate the uncertainty in $L_{\rm CO}$. The image rms is still used to define the 3$\sigma$ contours for CO emission.

The final sample of dust cores can be further refined based on their respective CO line profiles. The presence of multiple peaks within a line profile indicates that there are multiple sources along the line of sight contributing to the emission in the moment~0 maps and dust continuum images. In these cases, we cannot determine the fraction of dust emission corresponding to each source, and therefore the measured dust properties may not reflect an individual GMC. For this reason, we define a subsample {of} dust cores with approximately single-Gaussian $^{12}\rm CO$ and $^{13}\rm CO$ line profiles (hereafter referred to as `1G,' following the terminology of \citealp{Lada_2024}). This criterion also {excludes} 12 dust cores for which the $^{13}\rm CO$ peak is not present in the spectrum. As $^{13}\rm CO$ is expected to spatially align with the dust emission (see Section~\ref{Sec:maps}; \citealp{Forbrich_2020, Viaene_2021}), this further refines our sample to the highest-quality data to ensure reliable measurements. The 1G sample excludes 26 sources (5 resolved) from the full sample, including 15 from the \citet{Viaene_2021} sample.

\begin{figure}[h]
    \raggedleft
    \subfigure{\includegraphics[width=0.98\linewidth]{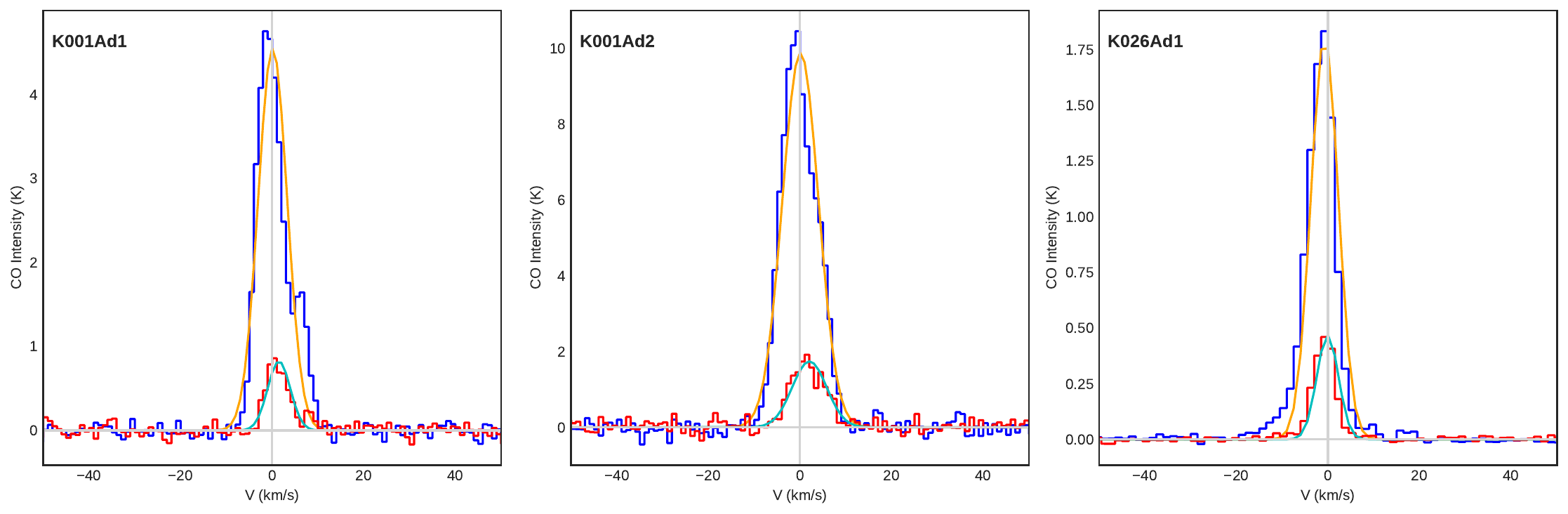}}
    \vspace{1em}
    \subfigure{\includegraphics[width=\linewidth]{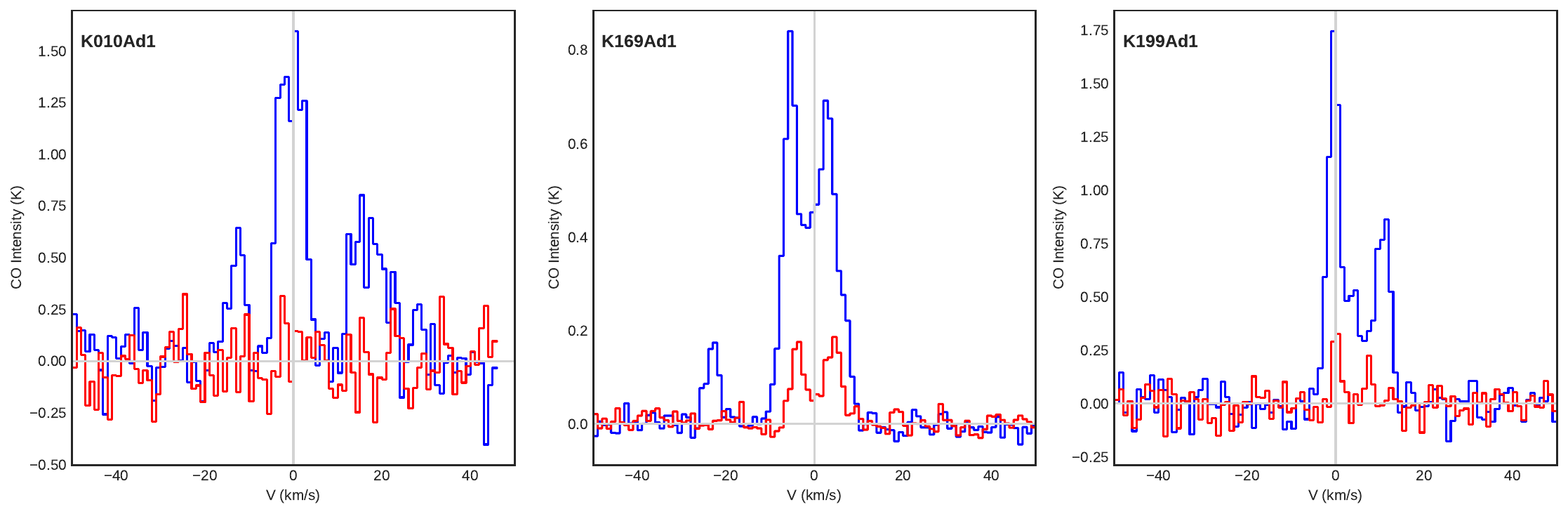}}
    \caption{CO line profiles for six individual dust cores. {The top panels show} three dust cores from the 1G sample, while {the bottom panels show} three dust cores with multiple-component CO line profiles. Both $^{12}\rm CO$ and $^{13}\rm CO$ {spectral lines} are displayed with the corresponding Gaussian fits.
}
    \label{fig:Spectra}
\end{figure}

In Figure~\ref{fig:Spectra}, we show CO line profiles for 1) three dust cores from our 1G sample and 2) three dust cores with multiple-component line profiles. We display both $^{12}\rm CO$ and $^{13}\rm CO$ line profiles with corresponding Gaussian fits for the 1G sources, from which the CO intensity was measured. The line profiles for the entire dust core sample are displayed in Figure~\ref{fig: Atlas}{,} along with the corresponding dust continuum images. CO line profiles extracted from the 1G sources are well represented by a single Gaussian fit. \citet{Lada_2024} calculated formal uncertainties in velocity dispersions and integrated intensities for CO line profiles extracted from this same survey, corresponding to {entire GMCs rather than dust cores, as is done} in this work. For $^{12}\rm CO$, the uncertainties in measurements of line dispersion and intensity are 3\% and 8\%, respectively. For $^{13}\rm CO$, uncertainties of 4\% and 15\% were found. These relatively small uncertainties highlight the high quality of these spectra. Therefore, we conclude that the uncertainties in these Gaussian fits are {likely} similar.

\section{Results and Discussion} \label{sec:Res}

The following section presents the results of our analysis of the dust core sample. After applying the methodology to this sample of 71 dust cores, we obtain {the} size, continuum flux, and $L_{\rm CO}$ for individual detections. We first combine $M_{\rm dust}$ with the corresponding CO(2-1) emission to measure $\alpha^{\prime}_{\rm CO}$ {(i.e., $M_{\rm dust}/L_{\rm CO}$)} for both $\rm ^{12}CO$ and $\rm ^{13}CO$ for all dust cores. For resolved dust cores, we perform a virial analysis to determine whether these are self-bound by gravity and approximately satisfy the virial theorem. Finally, we compare {our results} to H~{\sc ii} region metallicities from \citet{Bosomworth_2025} to update the direct test on the metallicity dependence of $\alpha^{\prime}_{\rm CO}$.

\subsection{Dust Continuum Images and CO Line Profiles}
\label{Sec:maps}

\begin{figure}[h]
    \centering
    \includegraphics[width=\linewidth]{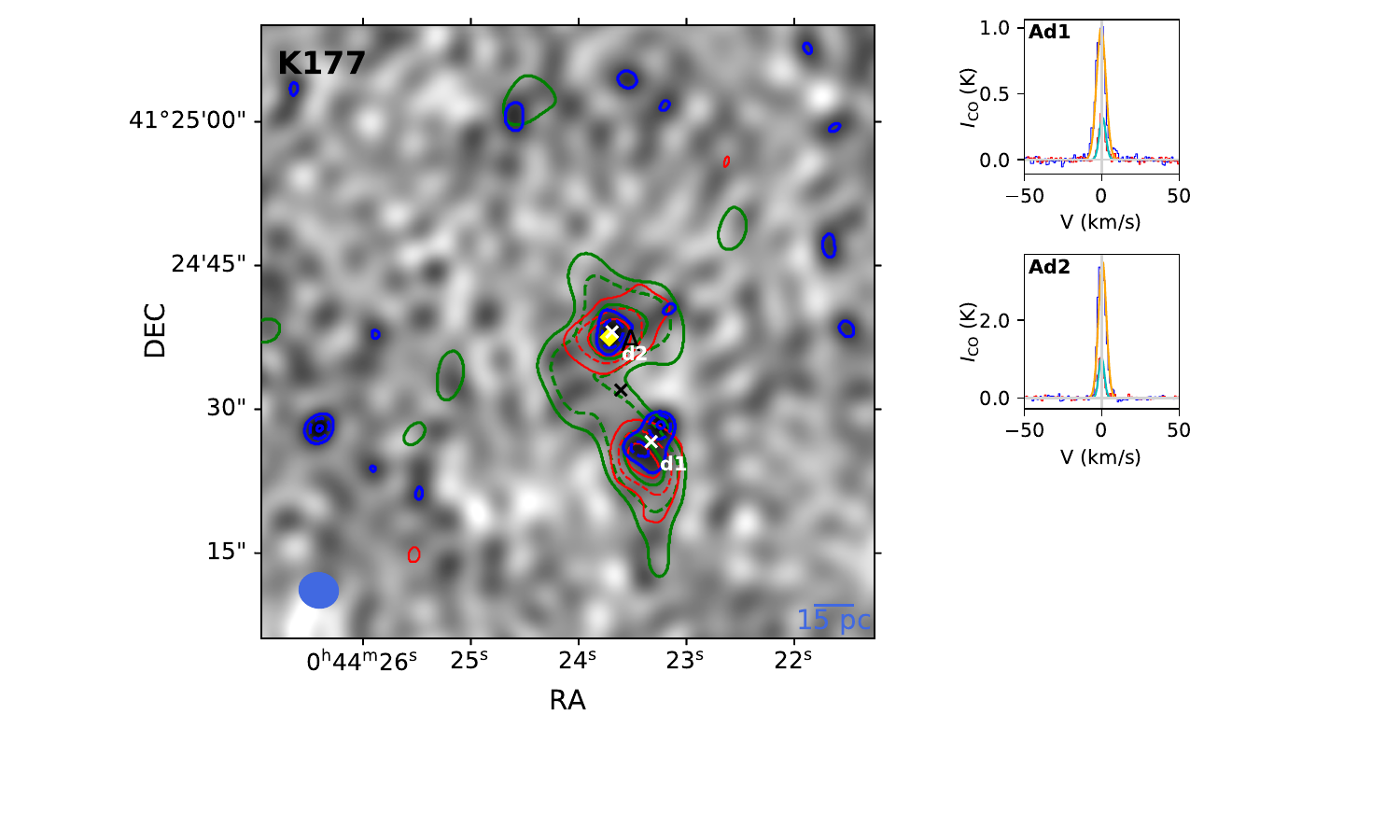}
    \caption{\textit{Left,} SMA maps of dust continuum of individual \citep{Kirk_2015} GMAs in M31, for fields which contain one or more dust cores. Blue contours display dust continuum emission at 2.5$\sigma$, 3.5$\sigma$ and 4.5$\sigma$, where $\sigma$ is the continuum image background rms, the values given in Table \ref{tab:props}. Green contours display $\rm ^{12}CO$ at 3, 6, 12, 24 and 48$\sigma$. Red contours display $\rm ^{13}CO$ at 3, 6 and 10$\sigma$. {The corresponding center of mass (by area) of individual GMCs as traced by $\rm ^{12}CO$ at 3$\sigma$ are marked by black `x's', and by white `x's' for individual dust cores as traced by dust continuum at 2.5$\sigma$.} \textit{Right,} Corresponding CO line profiles extracted from within individual dust cores, for both $\rm ^{12}CO$ and $\rm ^{13}CO$. Hectospec fibre positions corresponding to H~{\sc ii} regions identified from optical spectroscopy \citep{Bosomworth_2025} and associated with GMCs hosting a 1G dust core are shown as yellow diamonds (see Section \ref{Sec: met}). Gaussian fits to line profiles are displayed for the 1G sample. The complete figure set (51 images) is available in the online journal.
}
    \label{fig: Atlas}
\end{figure}

Dust continuum images of the 51 observed GMAs containing one or more dust cores are displayed in Figure~\ref{fig: Atlas}, along with the CO line profiles of any associated dust cores. The GMAs without an associated dust core are displayed in Appendix~\ref{Sec:NonDet}. CO {contours} in Figure~\ref{fig: Atlas} correspond to the CO(2-1) moment~0 maps, which have identical astrometry and ($u, v$) coverage to the continuum images, {allowing} us to directly compare their spatial extent.

In all maps, $^{12}\rm CO$ emission is more spatially extended than the $^{13}\rm CO$ emission, which is in turn more extended than the continuum emission. This indicates that, at the sensitivity of our observations, we {detect} $^{13}\rm CO$ and dust continuum emission from the higher column density regions of GMCs, while $^{12}\rm CO$ traces the entire extent of the molecular cloud. {NOEMA images presented by \citet{Forbrich_2023} show that the dust continuum emission in these observations is spatially similar to HCN and HCO$^{+}$, both of which have high dipole moments compared to CO and are assumed to trace dense gas (i.e. $n(\rm H_{2})~\gtrsim~10^{4}~\rm cm^{-3}$), indicating that we are probing similarly dense material at the available sensitivity.}

\subsection{Dust Core Properties}
\label{Sec:DCP}

\begin{figure}[h]
    \centering
    \begin{subfigure}
        \centering
        \includegraphics[width=0.45\linewidth]{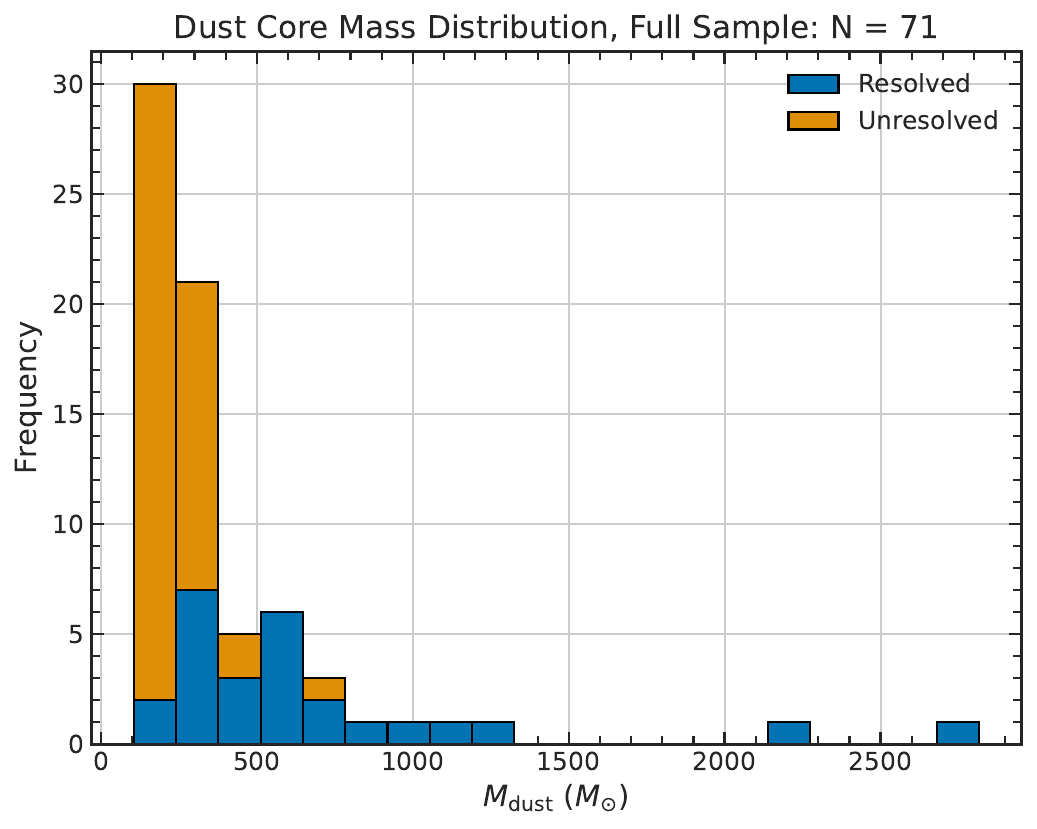}
    \end{subfigure}
    \begin{subfigure}
        \centering
        \includegraphics[width=0.45\linewidth]{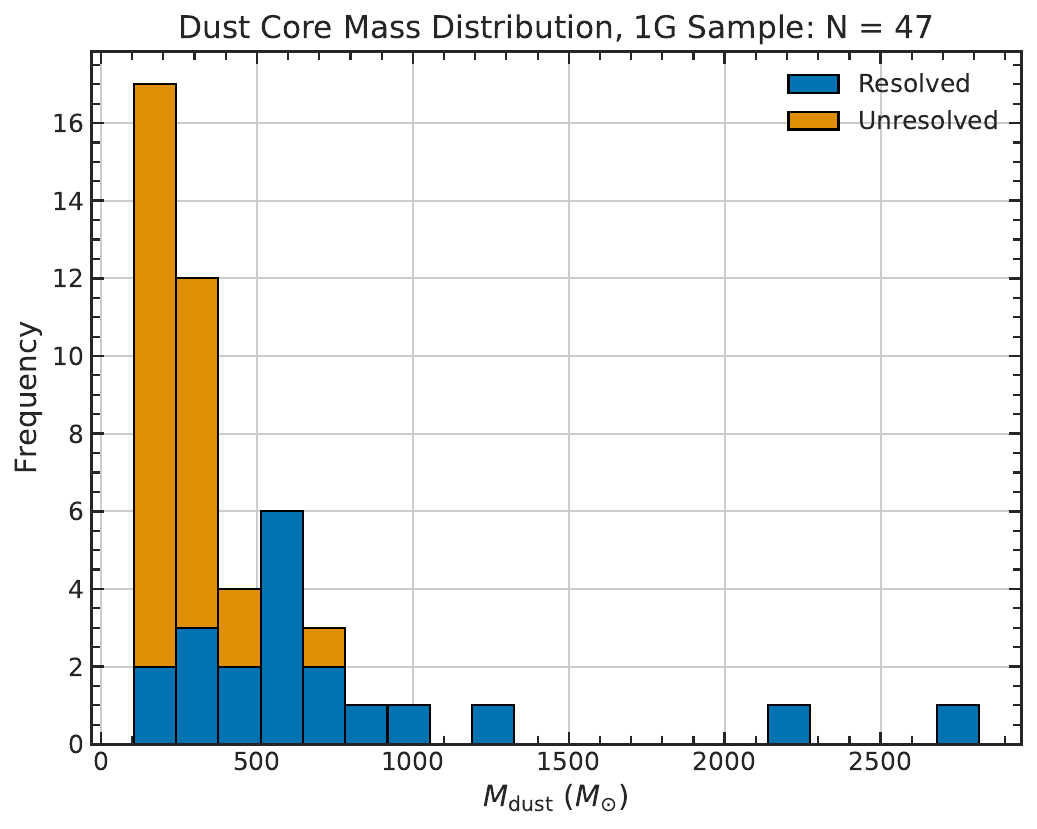}
    \end{subfigure}
    \caption{Histogram of dust masses for \textit{left,} our full sample of dust cores, of which 26 are resolved and 45 unresolved and \textit{right,} our 1G sample of dust cores, of which 20 are resolved and 27 unresolved.
    }
\label{fig:DustHist}
\end{figure}

The left panel of Figure \ref{fig:DustHist} displays the dust mass distribution for the entire sample of 71 dust cores, and the right panel shows the dust mass distribution for the 47 1G dust cores only, with the numbers of resolved and unresolved cores displayed separately. These include dust cores previously reported by \citet{Viaene_2021}, {but with updated dust masses calculated using the 2.5$\sigma$ detection threshold, as described in Section \ref{sec:DustMass}.} Resolved dust cores lie primarily in the dust mass range of $M_{\rm dust} \sim 200$--$700~M_{\odot}$. {The multi-component (hereafter `multi-G') sample contains only five resolved sources.} By comparing the two histograms in Figure \ref{fig:DustHist}, we conclude that the majority of the non-1G sources are unresolved and have $M_{\rm dust} \lesssim 200~M_{\odot}$. The remaining dust cores in our sample are resolved, and three have masses of $M_{\rm dust} > 1000~M_{\odot}$. {From lowest to highest mass, these are identified as K026Ad1, K191Ad1, and K213Ad1.}

We present the properties of individual dust cores from our final sample in Table~\ref{tab:props}, {derived as described above.} Dust core nomenclature is based on the corresponding \citet{Kirk_2015} GMA, {ordered alphabetically from highest to lowest CO mass of the individual GMC associations.} We then introduce d1, d2, etc. labels for dust cores, starting with the highest $M_{\rm dust}$ detection associated with a particular GMC. The deconvolved radii of our resolved dust cores range from $\sim 10$--23~pc, {while} the deconvolved radii of the 3$\sigma$ $\rm ^{12}CO$ contours range from $\sim 26$--61~pc, confirming that we do not trace the entire GMCs with dust emission.

\clearpage

\setlength{\tabcolsep}{3pt}
\setlength{\LTcapwidth}{\textwidth}
    
\begin{center}
\begin{longtable}{ccccccccc}
\caption{Table of dust core properties.}
\label{tab:props} \\
\hline
Source & RA, Dec.$^{a}$ & Radius (deconv.) & $F_{\rm cont}$ & rms & Beam Size & $M_{\rm dust}$ & $I_{\rm ^{12}CO}$ & $I_{\rm ^{13}CO}$ \\
& J2000 &~pc & mJy & mJy/beam & $\rm a \times b~(arcsec)$ & $\rm M_{\odot}$ & $\rm K~km~s^{-1}$ & $\rm K~km~s^{-1}$ \\
\hline
\endfirsthead
\caption{(continued).} \\
\hline
Source & RA, Dec. & Radius (deconv.) & $F_{\rm cont}$ & rms & Beam Size & $M_{\rm dust}$ & $I_{\rm ^{12}CO}$ & $I_{\rm ^{13}CO}$ \\
& J2000 &~pc & mJy & mJy/beam & $\rm a \times b~(arcsec)$ & $\rm M_{\odot}$ & $\rm K~km~s^{-1}$ & $\rm K~km~s^{-1}$ \\
\hline
\endhead

\hline
\endfoot

\hline
\endlastfoot
K001Ad1 & 00:39:10.2, +40:37:22.2 & 7.5 & 0.9 $~\pm~$ 0.2 & 0.2 & 4.4 $~\times~$ 3.4 & 280 $~\pm~$ 66 & 28.5 $~\pm~$ 0.2 & 4.0 $~\pm~$ 0.2 \\ K001Ad2 & 00:39:09.8, +40:37:16.8 & 4.8 & 0.6 $~\pm~$ 0.2 & 0.2 & 4.4 $~\times~$ 3.4 & 207 $~\pm~$ 66 & 34.9 $~\pm~$ 0.3 & 5.8 $~\pm~$ 0.2 \\ K008Ad1 & 00:42:27.2, +41:18:18.8 & 7.2 & 0.5 $~\pm~$ 0.1 & 0.1 & 4.1 $~\times~$ 3.6 & 170 $~\pm~$ 40 & 8.7 $~\pm~$ 0.5 & 0.2 $~\pm~$ 0.1 \\ K008Ad2 & 00:42:28.8, +41:18:22.3 & 4.3 & 0.3 $~\pm~$ 0.1 & 0.1 & 4.1 $~\times~$ 3.6 & 115 $~\pm~$ 40 & 30.5 $~\pm~$ 0.3 & 0.2 $~\pm~$ 0.0 \\ K010Ad1 & 00:42:34.5, +41:21:09.0 & 4.6 & 0.5 $~\pm~$ 0.1 & 0.1 & 3.9 $~\times~$ 3.4 & 158 $~\pm~$ 45 & 4.1 $~\pm~$ 0.3 & 0.2 $~\pm~$ 0.1 \\ K026Ad1* & 00:41:29.7, +41:04:56.2 & 20.8 & 4.0 $~\pm~$ 0.4 & 0.2 & 4.5 $~\times~$ 4.0 & 1315 $~\pm~$ 125 & 13.5 $~\pm~$ 0.1 & 2.8 $~\pm~$ 0.0 \\ K029Ad1 & 00:42:20.5, +41:16:02.5 & 5.1 & 0.5 $~\pm~$ 0.2 & 0.2 & 4.1 $~\times~$ 3.5 & 177 $~\pm~$ 51 & 16.4 $~\pm~$ 0.6 & 0.8 $~\pm~$ 0.2 \\ K060Ad1 & 00:41:00.1, +40:36:59.2 & 5.4 & 0.6 $~\pm~$ 0.2 & 0.2 & 4.2 $~\times~$ 3.5 & 183 $~\pm~$ 60 & 7.7 $~\pm~$ 0.2 & 1.4 $~\pm~$ 0.2 \\ K060Ad2 & 00:40:59.5, +40:36:45.5 & 4.6 & 0.5 $~\pm~$ 0.2 & 0.2 & 4.2 $~\times~$ 3.5 & 154 $~\pm~$ 60 & 1.4 $~\pm~$ 0.2 & 0.2 $~\pm~$ 0.1 \\ K063Ad1* & 00:40:40.1, +40:35:40.5 & 14.1 & 1.0 $~\pm~$ 0.2 & 0.1 & 3.8 $~\times~$ 3.5 & 325 $~\pm~$ 54 & 6.7 $~\pm~$ 0.1 & 0.9 $~\pm~$ 0.1 \\ K063Ad2 & 00:40:39.2, +40:35:38.9 & 4.6 & 0.4 $~\pm~$ 0.1 & 0.1 & 3.8 $~\times~$ 3.5 & 118 $~\pm~$ 32 & 8.0 $~\pm~$ 0.2 & 1.0 $~\pm~$ 0.2 \\ K063Ad3 & 00:40:39.0, +40:35:31.0 & 4.7 & 0.3 $~\pm~$ 0.1 & 0.1 & 3.8 $~\times~$ 3.5 & 105 $~\pm~$ 32 & 12.2 $~\pm~$ 0.3 & 1.7 $~\pm~$ 0.2 \\ K067Ad1* & 00:41:38.1, +40:43:56.6 & 10.1 & 0.7 $~\pm~$ 0.2 & 0.2 & 4.2 $~\times~$ 3.7 & 213 $~\pm~$ 60 & 11.0 $~\pm~$ 0.1 & 2.2 $~\pm~$ 0.1 \\ K071Ad1* & 00:41:00.8, +41:03:59.4 & 12.6 & 1.9 $~\pm~$ 0.3 & 0.2 & 4.0 $~\times~$ 3.6 & 627 $~\pm~$ 115 & 11.3 $~\pm~$ 0.2 & 2.4 $~\pm~$ 0.1 \\ K071Ad2 & 00:41:00.8, +41:03:36.4 & 4.3 & 0.8 $~\pm~$ 0.2 & 0.2 & 4.0 $~\times~$ 3.6 & 267 $~\pm~$ 78 & 23.7 $~\pm~$ 1.3 & 3.1 $~\pm~$ 0.2 \\ K078Bd1* & 00:41:28.0, +41:12:06.9 & 10.1 & 0.8 $~\pm~$ 0.2 & 0.2 & 4.0 $~\times~$ 3.4 & 248 $~\pm~$ 71 & 4.4 $~\pm~$ 0.2 & 0.4 $~\pm~$ 0.1 \\ K081Ad1 & 00:41:47.4, +41:18:55.2 & 6.9 & 0.7 $~\pm~$ 0.2 & 0.2 & 4.0 $~\times~$ 3.5 & 232 $~\pm~$ 66 & 18.5 $~\pm~$ 0.3 & 3.7 $~\pm~$ 0.2 \\ K092Ad1* & 00:41:22.1, +40:49:52.2 & 18.0 & 1.6 $~\pm~$ 0.2 & 0.2 & 6.9 $~\times~$ 5.0 & 522 $~\pm~$ 76 & 5.4 $~\pm~$ 0.1 & 1.4 $~\pm~$ 0.0 \\ K093Ad1* & 00:41:25.7, +40:49:55.9 & 11.7 & 1.3 $~\pm~$ 0.3 & 0.2 & 4.2 $~\times~$ 3.6 & 427 $~\pm~$ 96 & 9.2 $~\pm~$ 0.1 & 2.2 $~\pm~$ 0.4 \\ K094Ad1 & 00:41:29.2, +40:50:49.6 & 6.3 & 0.5 $~\pm~$ 0.2 & 0.2 & 4.2 $~\times~$ 3.3 & 166 $~\pm~$ 56 & 19.9 $~\pm~$ 0.2 & 3.5 $~\pm~$ 0.1 \\ K098Ad1 & 00:43:02.8, +41:37:21.5 & 5.0 & 0.7 $~\pm~$ 0.2 & 0.2 & 4.0 $~\times~$ 3.6 & 223 $~\pm~$ 80 & 19.0 $~\pm~$ 0.3 & 2.5 $~\pm~$ 0.2 \\ K098Bd1* & 00:42:59.5, +41:37:07.3 & 13.4 & 2.3 $~\pm~$ 0.4 & 0.2 & 4.0 $~\times~$ 3.6 & 746 $~\pm~$ 125 & 27.4 $~\pm~$ 0.2 & 5.3 $~\pm~$ 0.2 \\ K119Bd1* & 00:43:34.4, +41:09:43.9 & 18.4 & 3.1 $~\pm~$ 0.3 & 0.2 & 3.9 $~\times~$ 3.5 & 1002 $~\pm~$ 97 & 8.0 $~\pm~$ 0.1 & 1.7 $~\pm~$ 0.0 \\ K120Ad1* & 00:43:55.1, +41:12:06.5 & 10.2 & 0.9 $~\pm~$ 0.2 & 0.2 & 4.0 $~\times~$ 3.4 & 280 $~\pm~$ 73 & 5.0 $~\pm~$ 0.1 & 1.1 $~\pm~$ 0.1 \\ K121Dd1 & 00:42:43.3, +40:59:40.5 & 5.2 & 0.4 $~\pm~$ 0.1 & 0.1 & 3.6 $~\times~$ 3.2 & 127 $~\pm~$ 38 & 1.1 $~\pm~$ 0.2 & 0.3 $~\pm~$ 0.0 \\ K134Ad1 & 00:44:23.0, +41:49:18.0 & 8.2 & 0.9 $~\pm~$ 0.3 & 0.3 & 5.5 $~\times~$ 4.7 & 299 $~\pm~$ 92 & 8.6 $~\pm~$ 0.2 & 1.7 $~\pm~$ 0.1 \\ K134Ad2 & 00:44:23.8, +41:49:28.5 & 5.6 & 0.8 $~\pm~$ 0.3 & 0.3 & 5.5 $~\times~$ 4.7 & 250 $~\pm~$ 92 & 6.0 $~\pm~$ 0.1 & 0.7 $~\pm~$ 0.1 \\ K136Ad1* & 00:44:29.6, +41:51:43.8 & 21.7 & 3.3 $~\pm~$ 0.3 & 0.2 & 4.4 $~\times~$ 3.9 & 1086 $~\pm~$ 97 & 10.3 $~\pm~$ 0.4 & 2.3 $~\pm~$ 0.1 \\ K136Ad2 & 00:44:31.3, +41:51:53.1 & 4.3 & 0.5 $~\pm~$ 0.1 & 0.2 & 4.4 $~\times~$ 3.9 & 173 $~\pm~$ 42 & 8.6 $~\pm~$ 0.3 & 1.6 $~\pm~$ 0.1 \\ K138Ad1* & 00:44:36.9, +41:51:42.6 & 13.5 & 2.0 $~\pm~$ 0.2 & 0.1 & 3.5 $~\times~$ 3.3 & 641 $~\pm~$ 69 & 20.0 $~\pm~$ 0.1 & 4.0 $~\pm~$ 0.1 \\ K142Ad1 & 00:44:57.8, +41:55:24.4 & 8.6 & 0.8 $~\pm~$ 0.2 & 0.2 & 8.1 $~\times~$ 5.0 & 259 $~\pm~$ 79 & 0.6 $~\pm~$ 0.1 & 0.0 $~\pm~$ 0.0 \\ K142Ad2 & 00:44:56.9, +41:55:20.6 & 6.6 & 0.8 $~\pm~$ 0.2 & 0.2 & 8.1 $~\times~$ 5.0 & 246 $~\pm~$ 79 & 0.6 $~\pm~$ 0.1 & 0.0 $~\pm~$ 0.0 \\ K143Ad1 & 00:44:51.2, +41:54:38.7 & 4.8 & 0.7 $~\pm~$ 0.2 & 0.2 & 4.0 $~\times~$ 3.5 & 231 $~\pm~$ 67 & 9.9 $~\pm~$ 0.2 & 0.9 $~\pm~$ 0.2 \\ K149Ad1 & 00:45:36.6, +41:51:05.1 & 8.5 & 0.6 $~\pm~$ 0.2 & 0.2 & 3.9 $~\times~$ 3.5 & 178 $~\pm~$ 57 & 11.1 $~\pm~$ 0.2 & 2.0 $~\pm~$ 0.1 \\ K151Ad1 & 00:45:27.0, +41:44:26.8 & 7.6 & 0.6 $~\pm~$ 0.2 & 0.2 & 8.1 $~\times~$ 5.0 & 200 $~\pm~$ 68 & 8.4 $~\pm~$ 0.1 & 0.7 $~\pm~$ 0.1 \\ K153Ad1 & 00:45:27.5, +41:46:25.8 & 8.8 & 0.8 $~\pm~$ 0.2 & 0.2 & 8.0 $~\times~$ 4.9 & 253 $~\pm~$ 66 & 11.9 $~\pm~$ 0.1 & 1.4 $~\pm~$ 0.0 \\ K153Ad2 & 00:45:27.7, +41:46:38.9 & 5.3 & 0.6 $~\pm~$ 0.2 & 0.2 & 8.0 $~\times~$ 4.9 & 182 $~\pm~$ 66 & 9.2 $~\pm~$ 0.1 & 1.0 $~\pm~$ 0.0 \\ K154Ad1* & 00:45:36.0, +41:53:00.4 & 13.4 & 1.1 $~\pm~$ 0.2 & 0.1 & 4.0 $~\times~$ 3.7 & 347 $~\pm~$ 67 & 5.1 $~\pm~$ 0.1 & 0.8 $~\pm~$ 0.0 \\ K157Ad1* & 00:45:43.7, +41:52:43.9 & 7.5 & 0.9 $~\pm~$ 0.2 & 0.2 & 4.2 $~\times~$ 3.7 & 280 $~\pm~$ 58 & 11.7 $~\pm~$ 0.4 & 1.9 $~\pm~$ 0.1 \\ K157Bd1 & 00:45:41.8, +41:52:45.3 & 11.1 & 1.0 $~\pm~$ 0.2 & 0.2 & 4.2 $~\times~$ 3.7 & 321 $~\pm~$ 74 & 6.9 $~\pm~$ 0.2 & 0.2 $~\pm~$ 0.1 \\ K157Bd2 & 00:45:44.8, +41:52:40.8 & 4.9 & 0.6 $~\pm~$ 0.2 & 0.2 & 4.2 $~\times~$ 3.7 & 202 $~\pm~$ 58 & 7.6 $~\pm~$ 0.1 & 1.1 $~\pm~$ 0.1 \\ K160Ad1 & 00:45:07.7, +41:34:20.7 & 10.8 & 1.0 $~\pm~$ 0.2 & 0.2 & 8.1 $~\times~$ 5.0 & 324 $~\pm~$ 61 & 6.1 $~\pm~$ 0.1 & 1.0 $~\pm~$ 0.2 \\ K160Bd1 & 00:45:09.2, +41:34:33.1 & 6.5 & 0.7 $~\pm~$ 0.2 & 0.2 & 8.1 $~\times~$ 5.0 & 229 $~\pm~$ 61 & 4.0 $~\pm~$ 0.3 & 0.5 $~\pm~$ 0.1 \\ K160Cd1 & 00:45:08.6, +41:34:52.8 & 14.6 & 1.2 $~\pm~$ 0.2 & 0.2 & 8.1 $~\times~$ 5.0 & 389 $~\pm~$ 61 & 5.2 $~\pm~$ 0.6 & 0.6 $~\pm~$ 0.1 \\ K162Ad1* & 00:45:11.5, +41:36:58.0 & 14.3 & 1.4 $~\pm~$ 0.2 & 0.2 & 5.2 $~\times~$ 4.3 & 463 $~\pm~$ 77 & 23.1 $~\pm~$ 0.5 & 3.6 $~\pm~$ 0.6 \\ K162Ad2 & 00:45:10.8, +41:37:08.0 & 6.7 & 0.6 $~\pm~$ 0.2 & 0.2 & 5.2 $~\times~$ 4.3 & 208 $~\pm~$ 58 & 16.6 $~\pm~$ 0.2 & 2.3 $~\pm~$ 0.4 \\ K169Ad1* & 00:44:38.2, +41:25:10.9 & 13.6 & 1.3 $~\pm~$ 0.2 & 0.1 & 3.6 $~\times~$ 3.3 & 434 $~\pm~$ 65 & 8.8 $~\pm~$ 0.2 & 1.7 $~\pm~$ 0.1 \\ K170Ad1* & 00:44:42.5, +41:27:36.8 & 15.2 & 1.7 $~\pm~$ 0.3 & 0.1 & 3.5 $~\times~$ 3.2 & 562 $~\pm~$ 84 & 15.2 $~\pm~$ 0.3 & 2.9 $~\pm~$ 0.1 \\ K171Ad1 & 00:44:55.7, +41:29:14.7 & 7.4 & 0.8 $~\pm~$ 0.2 & 0.2 & 4.0 $~\times~$ 3.6 & 252 $~\pm~$ 69 & 15.5 $~\pm~$ 0.3 & 3.7 $~\pm~$ 0.2 \\ K174Ad1 & 00:44:24.2, +41:21:04.8 & 7.5 & 0.9 $~\pm~$ 0.2 & 0.2 & 3.9 $~\times~$ 3.5 & 289 $~\pm~$ 56 & 6.7 $~\pm~$ 0.4 & 1.2 $~\pm~$ 0.1 \\ K174Cd1 & 00:44:23.8, +41:20:29.2 & 4.8 & 0.7 $~\pm~$ 0.2 & 0.2 & 3.9 $~\times~$ 3.5 & 212 $~\pm~$ 56 & 4.0 $~\pm~$ 0.2 & 0.2 $~\pm~$ 0.1 \\ K176Ad1* & 00:45:00.2, +41:28:32.5 & 19.3 & 1.9 $~\pm~$ 0.3 & 0.2 & 8.1 $~\times~$ 5.0 & 624 $~\pm~$ 94 & 10.4 $~\pm~$ 0.0 & 1.8 $~\pm~$ 0.3 \\ K177Ad1* & 00:44:25.0, +41:24:47.9 & 9.8 & 0.4 $~\pm~$ 0.1 & 0.1 & 3.8 $~\times~$ 3.4 & 141 $~\pm~$ 35 & 7.5 $~\pm~$ 0.1 & 1.7 $~\pm~$ 0.0 \\ K177Ad2 & 00:44:24.1, +41:24:42.8 & 7.2 & 0.4 $~\pm~$ 0.1 & 0.1 & 3.8 $~\times~$ 3.4 & 134 $~\pm~$ 29 & 18.5 $~\pm~$ 0.1 & 3.9 $~\pm~$ 0.1 \\ K190Ad1* & 00:44:30.3, +41:56:29.6 & 15.0 & 1.8 $~\pm~$ 0.2 & 0.2 & 4.9 $~\times~$ 4.2 & 598 $~\pm~$ 68 & 4.6 $~\pm~$ 0.1 & 0.4 $~\pm~$ 0.0 \\ K191Ad1* & 00:44:01.5, +41:49:11.2 & 16.0 & 6.9 $~\pm~$ 0.8 & 0.4 & 4.2 $~\times~$ 3.3 & 2270 $~\pm~$ 254 & 34.9 $~\pm~$ 0.4 & 6.6 $~\pm~$ 1.1 \\ K192Ad1 & 00:43:55.9, +41:48:22.5 & 4.9 & 1.3 $~\pm~$ 0.4 & 0.3 & 4.0 $~\times~$ 3.7 & 411 $~\pm~$ 120 & 20.0 $~\pm~$ 0.3 & 4.8 $~\pm~$ 0.3 \\ K199Ad1 & 00:43:42.0, +41:45:14.7 & 7.0 & 0.7 $~\pm~$ 0.2 & 0.2 & 4.0 $~\times~$ 3.6 & 233 $~\pm~$ 61 & 4.8 $~\pm~$ 0.4 & 0.5 $~\pm~$ 0.1 \\ K213Ad1* & 00:42:13.9, +40:51:17.2 & 22.9 & 8.6 $~\pm~$ 0.8 & 0.3 & 4.3 $~\times~$ 3.7 & 2816 $~\pm~$ 252 & 18.5 $~\pm~$ 0.1 & 3.3 $~\pm~$ 0.1 \\ K238Cd1 & 00:44:10.8, +41:35:08.4 & 5.7 & 0.7 $~\pm~$ 0.2 & 0.2 & 7.8 $~\times~$ 4.9 & 217 $~\pm~$ 79 & 4.3 $~\pm~$ 0.2 & 0.4 $~\pm~$ 0.1 \\ K239Ad1* & 00:44:12.0, +41:34:11.5 & 15.8 & 1.1 $~\pm~$ 0.2 & 0.2 & 7.8 $~\times~$ 4.9 & 368 $~\pm~$ 80 & 1.5 $~\pm~$ 0.0 & 0.2 $~\pm~$ 0.0 \\ K239Ad2 & 00:44:12.5, +41:34:07.3 & 3.4 & 0.7 $~\pm~$ 0.2 & 0.2 & 7.8 $~\times~$ 4.9 & 215 $~\pm~$ 70 & 10.6 $~\pm~$ 0.1 & 1.5 $~\pm~$ 0.2 \\ K240Ad1 & 00:44:11.8, +41:33:07.4 & 4.8 & 0.7 $~\pm~$ 0.2 & 0.2 & 4.0 $~\times~$ 3.5 & 241 $~\pm~$ 72 & 7.0 $~\pm~$ 0.2 & 1.0 $~\pm~$ 0.1 \\ K240Bd1 & 00:44:12.7, +41:33:13.8 & 4.9 & 0.7 $~\pm~$ 0.2 & 0.2 & 4.0 $~\times~$ 3.5 & 233 $~\pm~$ 72 & 7.7 $~\pm~$ 0.2 & 0.6 $~\pm~$ 0.2 \\ K270Ad1* & 00:39:18.4, +40:21:53.1 & 18.4 & 2.4 $~\pm~$ 0.2 & 0.1 & 3.5 $~\times~$ 3.2 & 787 $~\pm~$ 72 & 7.9 $~\pm~$ 0.0 & 1.2 $~\pm~$ 0.0 \\ K270Ad2 & 00:39:19.3, +40:22:00.8 & 4.6 & 0.3 $~\pm~$ 0.1 & 0.1 & 3.5 $~\times~$ 3.2 & 109 $~\pm~$ 29 & 10.9 $~\pm~$ 0.1 & 1.5 $~\pm~$ 0.1 \\ K273Bd1* & 00:39:43.2, +40:20:43.0 & 11.1 & 0.9 $~\pm~$ 0.2 & 0.2 & 4.0 $~\times~$ 3.5 & 295 $~\pm~$ 73 & 0.8 $~\pm~$ 0.2 & 0.1 $~\pm~$ 0.0 \\ K275Bd1 & 00:39:30.3, +40:21:09.8 & 5.4 & 0.8 $~\pm~$ 0.2 & 0.2 & 4.2 $~\times~$ 3.4 & 266 $~\pm~$ 68 & 9.5 $~\pm~$ 0.2 & 1.4 $~\pm~$ 0.1 \\ K291Ad1 & 00:46:33.3, +42:11:31.5 & 9.5 & 1.1 $~\pm~$ 0.2 & 0.2 & 4.4 $~\times~$ 3.8 & 358 $~\pm~$ 66 & 5.4 $~\pm~$ 0.1 & 0.6 $~\pm~$ 0.1 \\ K297Ad1 & 00:46:07.9, +42:11:26.0 & 15.0 & 2.1 $~\pm~$ 0.2 & 0.2 & 7.8 $~\times~$ 5.0 & 675 $~\pm~$ 62 & 12.6 $~\pm~$ 0.1 & 1.8 $~\pm~$ 0.1 \\ K301Ad1* & 00:41:19.3, +41:19:47.8 & 15.4 & 2.3 $~\pm~$ 0.2 & 0.1 & 3.5 $~\times~$ 3.2 & 755 $~\pm~$ 77 & 22.1 $~\pm~$ 0.1 & 4.2 $~\pm~$ 0.1 \\
\hline
\end{longtable}
\end{center}
\vspace{-3em}
\noindent\textbf{Notes.} $^{*}$ Resolved dust cores. \\
\noindent$^{a}$ Right Ascention (RA) and Declination (Dec.) gives coordinates of the centre of mass of the dust core.

\clearpage

Despite the identical astrometry, the peaks of CO and dust emission do not always spatially align. {One possible explanation, proposed by \citet{Viaene_2021}, is that dust peaks may correspond to hotter dust heated by ongoing star formation at the edges of GMCs, and are therefore spatially separated from the cold gas.} The possibility that continuum emission originates from an associated H~{\sc ii} region and is contaminated by free-free emission has already been investigated by \citet{Forbrich_2023}, {who found that at 230~GHz, free-free emission contributes only 6--23\% of the continuum flux for clouds with VLA-detected H~{\sc ii} regions.}

The relative sensitivity of our observations to different tracers is crucial for understanding the structure of molecular gas in GMCs. Since CO and continuum observations were obtained simultaneously with the same calibration, imaging, and astrometry, {experimental offsets are eliminated.} From the areas of the 2.5$\sigma$ contours for dust emission and the 3$\sigma$ contours for CO, we find that dust traces {10\%--60\% of the area traced by $\rm ^{12}CO$}, and 35\%--80\% of the area traced by $\rm ^{13}CO$ emission at the sensitivity achieved. {The filamentary nature of molecular clouds implies that at these smaller scales we are probing higher column densities \citep{Dame_2023}. Furthermore, dust and $\rm ^{13}CO$ emission from this SMA survey is cospatial with HCN from NOEMA \citep{Forbrich_2023}, indicating that dust emission is tracing high-volume-density gas, compared to the more diffuse gas traced by $\rm ^{12}CO$.}

HCN is an assumed dense molecular gas tracer, {tracing $\rm H_{2}$ densities of $\gtrsim 3 \times 10^{4}~\rm cm^{-3}$}, compared to CO, which traces densities of $\gtrsim 300~\rm cm^{-3}$ \citep{Gao_2004}. However, recent studies suggest that HCN can also be detected at lower densities \citep{Pety_2017, Dame_2023}. \citet{Forbrich_2023} found that the dust emission from the SMA discussed in this work and HCN from NOEMA observations for a subset of six M31 GMCs {are strongly spatially coincident}, suggesting that both surveys independently trace the high-column-density gas in the GMCs, {reflecting the limited sensitivity.} Our observations are therefore consistent with the idea that, at the sensitivity achieved in this study, dust can trace dense molecular gas in GMCs, which is closely related to the SFR (e.g., \citealp{Lada_2010}, \citeyear{Lada_2012}).

\subsection{CO Conversion Factors} \label{sec:alphamethod}

To convert CO luminosity to cloud mass, we require a conversion factor typically defined as $\alpha_{\rm CO} = M_{\rm tot}/L_{\rm CO}$, where $M_{\rm tot}$ is the total molecular mass of the cloud. {Knowledge of the CO conversion factor is critical when dust emission is not detected and only CO is available, such as in distant extragalactic sources.} Here, we use dust emission to calibrate $\alpha_{\rm CO}$. From our observations of dust and CO emission, we can directly measure the dust mass-to-light ratio, $\alpha^{\prime}_{\rm CO}$. {With an assumed $R_{\rm g/d}$, $\alpha^{\prime}_{\rm CO}$ can then be converted to $\alpha_{\rm CO}$ by $\alpha_{\rm CO} = \alpha^{\prime}_{\rm CO} \times R_{\rm g/d}$.}

For M31 GMCs, we calculate $\alpha^{\prime}_{\rm CO}$ for both $\rm ^{12}CO(2-1)$ and $\rm ^{13}CO(2-1)$, {denoted as} $\alpha^{\prime}_{\rm ^{12}CO}$ and $\alpha^{\prime}_{\rm ^{13}CO}$, {respectively}. Since in all cases both values correspond to identical spatial areas, this provides insight into the differences between the two isotopologues as molecular gas tracers. {Because $\rm ^{12}CO$ is the most easily observable and often the only tracer available for distant GMCs, calculating $\alpha^{\prime}_{\rm ^{13}CO}$ for the same sources allows us to assess potential optical depth effects in $\rm ^{12}CO$ by checking the consistency of the derived cloud dust masses.} 

\begin{figure}[ht]
    \centering
    \includegraphics[height = \linewidth, angle=90]{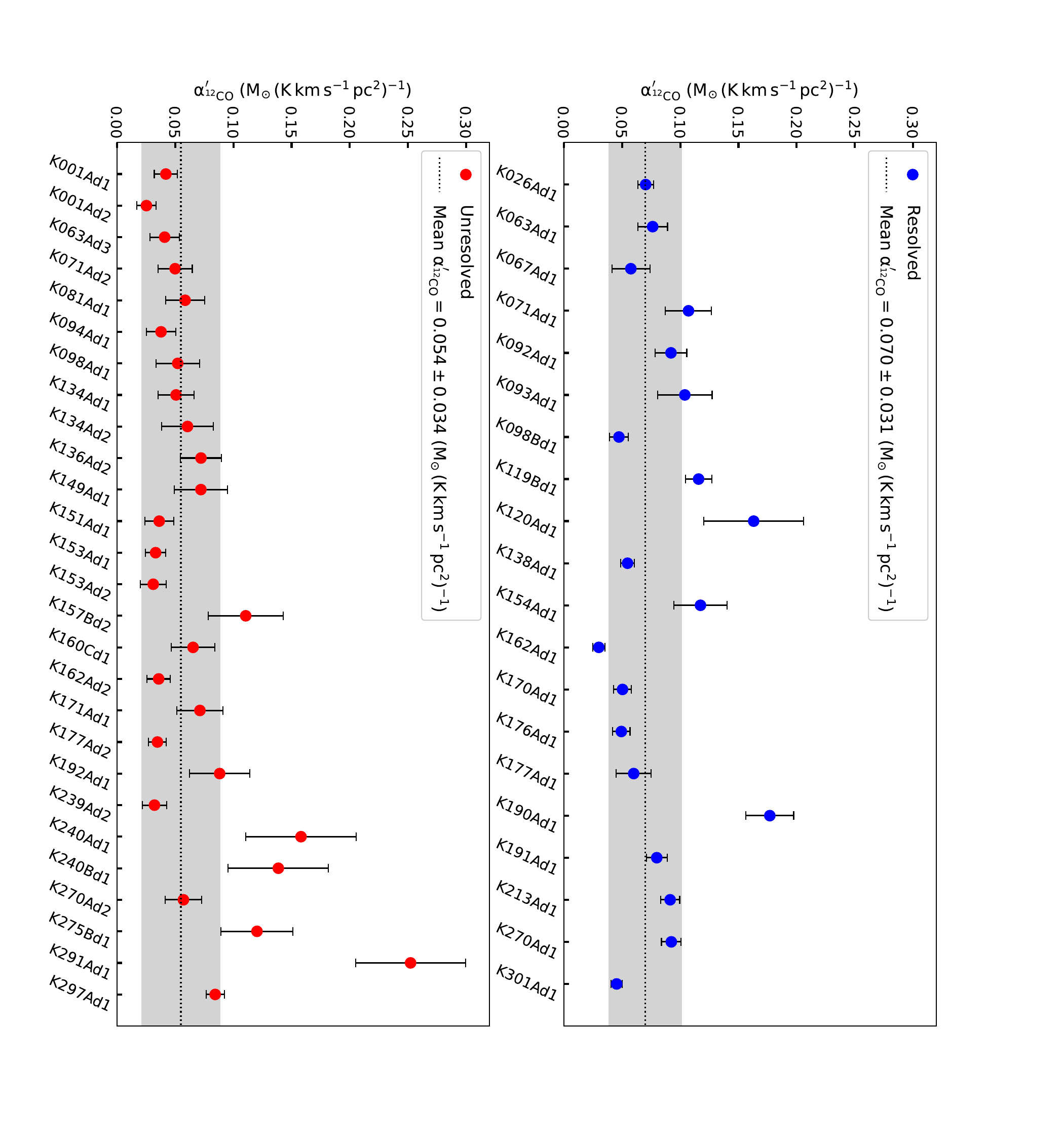}
    \caption{$^{12}\rm CO$ conversion factors, $\alpha^{\prime}_{^{12}\rm CO}$, for individual 1G dust cores for: \textit{top,} resolved sources and \textit{bottom,} unresolved sources. We display the mean $\alpha^{\prime}_{^{12}\rm CO}$ calculated for resolved and unresolved sources as a dotted line, with the 1$\sigma$ standard deviation represented by the shaded region.}
    \label{fig:alpha12}
\end{figure}

We compute $\alpha^{\prime}_{\rm CO}$ for the 1G dust core sample only, {as} for CO line profiles with multiple peaks, it is difficult to determine whether the dust emission originates from the GMC or from unrelated structures along the line of sight. Individual dust core $\alpha^{\prime}_{\rm ^{12}CO}$ values are displayed in Figure \ref{fig:alpha12} and reported in Table \ref{tab: sample}. {A minimum uncertainty of 0.01 is adopted.} The weighted mean value of $\alpha^{\prime}_{\rm ^{12}\rm CO}$ for resolved dust cores is $0.070 \pm 0.031~M_{\odot}\,(\rm K\, km\, s^{-1}\, pc^{2})^{-1}$, which is in excellent agreement with the previous measurement of $0.064 \pm 0.029~M_{\odot}\,(\rm K\, km\, s^{-1}\,pc^{2})^{-1}$ from \cite{Viaene_2021}. The mean $\alpha^{\prime}_{\rm ^{12}CO}$ values for resolved and unresolved dust cores, shown in Figure \ref{fig:alpha12}, {are consistent within $1\sigma$}. However, for unresolved dust cores, {the source size cannot be reliably determined, introducing additional uncertainty in the corresponding measurements.} The uncertainties reported in the mean $\alpha^{\prime}_{\rm CO}$ from all dust cores represent the standard deviation of the individual dust core measurements and are similar in magnitude to those found by \citet{Viaene_2021}. 

{Although no explicit $\alpha^{\prime}_{\rm ^{12}CO}$ value has been calculated for MW GMCs, one can estimate it from the \citet{Bolatto_2013} value of $\alpha_{\rm CO}$ for $^{12}\rm CO(1-0)$, assuming a gas-to-dust ratio of $R_{\rm g/d} \sim 136$ and an intensity ratio of $I_{\rm CO}(1-0)/I_{\rm CO}(2-1) = 0.7$. This yields a MW value of $\alpha^{\prime}_{\rm ^{12}CO} \sim 0.045~M_{\odot}\,(\rm K\,km\,s^{-1}\,pc^{2})^{-1}~\pm~0.3~dex$.} The value calculated here for M31 agrees within $<1\sigma$, {with a smaller relative uncertainty compared to the MW estimate.}

\begin{center}
\begin{longtable}{@{\extracolsep{\fill}} ccc}
    \caption{$\alpha^{\prime}_{\rm CO}$ for individual 1G dust cores from both $L_{\rm ^{12}CO}$ and $L_{\rm ^{13}CO}$.}
    \label{tab: sample} \\
    \hline
    Source & $\alpha^{\prime}_{\rm ^{12}CO}$ & $\alpha^{\prime}_{\rm ^{13}CO}$ \\
    & $\rm M_{\odot}~\rm (K~km~s^{-1}~pc^{2})^{-1}$ & $\rm M_{\odot}~\rm (K~km~s^{-1}~pc^{2})^{-1}$ \\
    \hline
    \endfirsthead
    \caption{(continued).} \\
    \hline
    Source & $\alpha^{\prime}_{\rm ^{12}CO}$ & $\alpha^{\prime}_{\rm ^{13}CO}$ \\
    & $\rm M_{\odot}~\rm (K~km~s^{-1}~pc^{2})^{-1}$ & $\rm M_{\odot}~\rm (K~km~s^{-1}~pc^{2})^{-1}$ \\
    \hline
    \endhead
    
    \hline
    \endfoot
    
    \hline
    \endlastfoot

    K001Ad1 & 0.04 $~\pm~$ 0.01 & 0.30 $~\pm~$ 0.07 \\ K001Ad2 & 0.03 $~\pm~$ 0.01 & 0.15 $~\pm~$ 0.05 \\ K026Ad1* & 0.07 $~\pm~$ 0.01 & 0.34 $~\pm~$ 0.03 \\ K063Ad1* & 0.08 $~\pm~$ 0.01 & 0.56 $~\pm~$ 0.12 \\ K063Ad3 & 0.04 $~\pm~$ 0.01 & 0.29 $~\pm~$ 0.10 \\ K067Ad1* & 0.06 $~\pm~$ 0.02 & 0.29 $~\pm~$ 0.08 \\ K071Ad1* & 0.11 $~\pm~$ 0.02 & 0.52 $~\pm~$ 0.10 \\ K071Ad2 & 0.05 $~\pm~$ 0.01 & 0.38 $~\pm~$ 0.11 \\ K081Ad1 & 0.06 $~\pm~$ 0.02 & 0.29 $~\pm~$ 0.09 \\ K092Ad1* & 0.09 $~\pm~$ 0.01 & 0.36 $~\pm~$ 0.05 \\ K093Ad1* & 0.10 $~\pm~$ 0.02 & 0.43 $~\pm~$ 0.12 \\ K094Ad1 & 0.04 $~\pm~$ 0.01 & 0.21 $~\pm~$ 0.07 \\ K098Ad1 & 0.05 $~\pm~$ 0.02 & 0.40 $~\pm~$ 0.15 \\ K098Bd1* & 0.05 $~\pm~$ 0.01 & 0.24 $~\pm~$ 0.04 \\ K119Bd1* & 0.12 $~\pm~$ 0.01 & 0.54 $~\pm~$ 0.05 \\ K120Ad1* & 0.16 $~\pm~$ 0.04 & 0.76 $~\pm~$ 0.21 \\ K134Ad1 & 0.05 $~\pm~$ 0.02 & 0.26 $~\pm~$ 0.08 \\ K134Ad2 & 0.06 $~\pm~$ 0.02 & 0.52 $~\pm~$ 0.20 \\ K136Ad2 & 0.07 $~\pm~$ 0.02 & 0.39 $~\pm~$ 0.10 \\ K138Ad1* & 0.05 $~\pm~$ 0.01 & 0.27 $~\pm~$ 0.03 \\ K149Ad1 & 0.07 $~\pm~$ 0.02 & 0.41 $~\pm~$ 0.13 \\ K151Ad1 & 0.04 $~\pm~$ 0.01 & 0.44 $~\pm~$ 0.16 \\ K153Ad1 & 0.03 $~\pm~$ 0.01 & 0.29 $~\pm~$ 0.08 \\ K153Ad2 & 0.03 $~\pm~$ 0.01 & 0.29 $~\pm~$ 0.11 \\ K154Ad1* & 0.12 $~\pm~$ 0.02 & 0.74 $~\pm~$ 0.15 \\ K157Bd2 & 0.11 $~\pm~$ 0.03 & 0.76 $~\pm~$ 0.24 \\ K160Cd1 & 0.07 $~\pm~$ 0.02 & 0.60 $~\pm~$ 0.20 \\ K162Ad1* & 0.03 $~\pm~$ 0.01 & 0.19 $~\pm~$ 0.04 \\ K162Ad2 & 0.04 $~\pm~$ 0.01 & 0.26 $~\pm~$ 0.08 \\ K170Ad1* & 0.05 $~\pm~$ 0.01 & 0.26 $~\pm~$ 0.04 \\ K171Ad1 & 0.07 $~\pm~$ 0.02 & 0.30 $~\pm~$ 0.08 \\ K176Ad1* & 0.05 $~\pm~$ 0.01 & 0.28 $~\pm~$ 0.06 \\ K177Ad1* & 0.06 $~\pm~$ 0.02 & 0.27 $~\pm~$ 0.07 \\ K177Ad2 & 0.03 $~\pm~$ 0.01 & 0.17 $~\pm~$ 0.04 \\ K190Ad1* & 0.18 $~\pm~$ 0.02 & 1.84 $~\pm~$ 0.28 \\ K191Ad1* & 0.08 $~\pm~$ 0.01 & 0.42 $~\pm~$ 0.08 \\ K192Ad1 & 0.09 $~\pm~$ 0.03 & 0.37 $~\pm~$ 0.11 \\ K213Ad1* & 0.09 $~\pm~$ 0.01 & 0.51 $~\pm~$ 0.05 \\ K239Ad2 & 0.03 $~\pm~$ 0.01 & 0.23 $~\pm~$ 0.09 \\ K240Ad1 & 0.16 $~\pm~$ 0.05 & 1.05 $~\pm~$ 0.34 \\ K240Bd1 & 0.14 $~\pm~$ 0.04 & 1.63 $~\pm~$ 0.62 \\ K270Ad1* & 0.09 $~\pm~$ 0.01 & 0.59 $~\pm~$ 0.06 \\ K270Ad2 & 0.06 $~\pm~$ 0.02 & 0.42 $~\pm~$ 0.12 \\ K275Bd1 & 0.12 $~\pm~$ 0.03 & 0.79 $~\pm~$ 0.21 \\ K291Ad1 & 0.25 $~\pm~$ 0.05 & 2.20 $~\pm~$ 0.56 \\ K297Ad1 & 0.08 $~\pm~$ 0.01 & 0.58 $~\pm~$ 0.06 \\ K301Ad1* & 0.05 $~\pm~$ 0.00 & 0.24 $~\pm~$ 0.03 \\
    \hline
\end{longtable}
\end{center}
\vspace{-4.5em}
\begin{center}
\begin{minipage}{0.73\textwidth}
    \textbf{Notes.} $^{*}$ Resolved dust cores. 
\end{minipage}
\end{center}
\clearpage

\begin{figure}[h]
    \centering
    \includegraphics[height = \linewidth, angle=90]{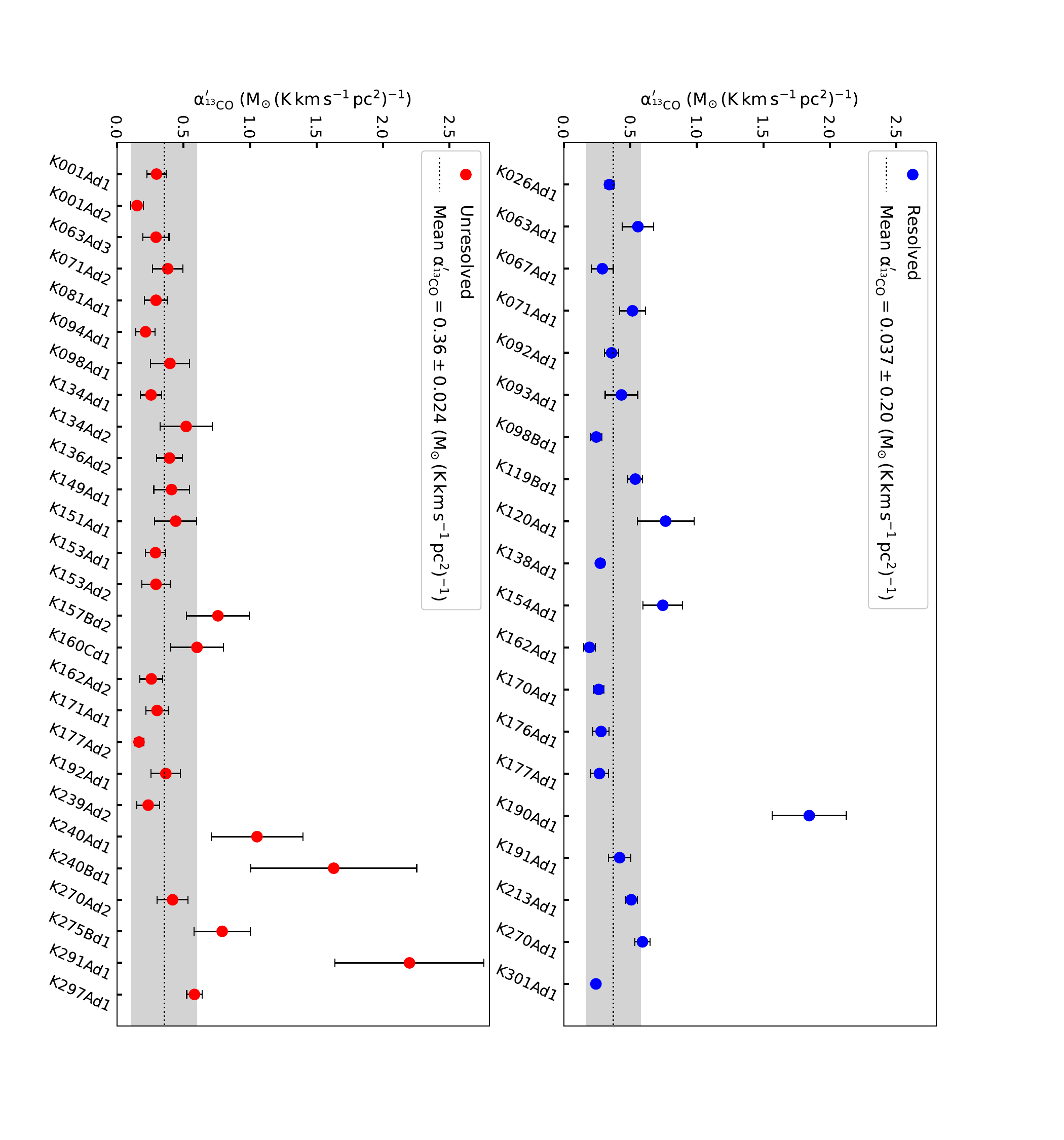}
    \caption{$^{13}\rm CO$ conversion factors, $\alpha^{\prime}_{^{13}\rm CO}$, for individual 1G dust clouds for: \textit{top,} resolved sources and \textit{bottom,} unresolved sources. As in Figure \ref{fig:alpha12}, we display the mean $\alpha^{\prime}_{^{13}\rm CO}$ calculated for resolved and unresolved sources as a dotted line, with the 1$\sigma$ standard deviation represented by the shaded region.}
    \label{fig:alpha13}
\end{figure}

As done for $\alpha^{\prime}_{\rm ^{12}CO}$, individual $\alpha^{\prime}_{\rm ^{13}\rm CO}$ values for 1G dust cores in M31 are displayed in Figure \ref{fig:alpha13} and reported in Table~\ref{tab: sample}. The mean $\alpha^{\prime}_{\rm ^{13}\rm CO}$ of resolved sources is 0.37 $\pm$ 0.20 $\rm M_{\odot}\,(\rm K\,km\,s^{-1}\,pc^{2})^{-1}$ which is in excellent agreement with the previous value of 0.36 $\pm$ 0.15 $\rm M_{\odot}\,(\rm K\,km\,s^{-1}\,pc^{2})^{-1}$ reported by \cite{Viaene_2021}. We see that the mean $\alpha^{\prime}_{\rm ^{13}\rm CO}$ for resolved and unresolved dust cores are again in agreement within 1$\sigma$.

In this study, we calculate $\alpha^{\prime}_{\rm CO}$ from {dust core emission} rather than from the entire GMC. The differences between values derived from dust cores {within the same GMC} are smaller than the overall standard deviation, {indicating that these measurements are representative of the entire GMCs within the uncertainties}. We examined $\alpha^{\prime}_{\rm CO}$ variations within the best-resolved dust cores. Only marginal nominal differences were found between pixels separated by more than one beam, and smaller-scale differences are likely temperature-driven rather than true $\alpha^{\prime}_{\rm CO}$ variations, especially near H\,{\sc ii} regions. We therefore find no compelling evidence for significant $\alpha^{\prime}_{\rm CO}$ variation within dust cores.

\citet{Lada_2024} found that M31 GMCs belong to one of two classes: dense and diffuse. Dense GMCs are identified by their strong $\rm ^{13}CO(2-1)$ emission lines. We {further} expect that all 1G dust cores are associated with dense GMCs, as all display measurable $\rm ^{13}CO(2-1)$ emission.

In dust cores K071Ad1, K119Bd1, K120Ad1, K154Ad1, and K190Ad1, the dust emission peak lies near the edge of the GMC, where CO emission is weaker. {Consequently,} values of $L_{\rm ^{12}CO}$ and sometimes $L_{\rm ^{13}CO}$ are lower than at the cloud centre, where the peak CO emission occurs. As a result, the ratio of $M_{\rm dust}/ L_{\rm CO}$ is biased high. This may be due to variations in $T_{\rm dust}$ within the GMC. For source K190Ad1 to have $\alpha^{\prime}_{\rm ^{12}\rm CO} = 0.07~M_{\odot}\,(\rm K\,km\,s^{-1}\,pc^{2})^{-1}$, a $T_{\rm dust}$ of 38 K would be required. The other aforementioned outliers would require $25~\text{K} \lesssim T_{\rm dust} \lesssim 36~\text{K}$. In Orion, $T_{\rm dust}$ derived from continuum emission is typically $20-30$~K (e.g., \citealp{Bouvier_2021}), but can be as high as $T_{\rm dust} \lesssim 50$ K in warmer regions \citep{Shirley_2005}. In M17, $T_{\rm dust}$ ranges from $\sim 10$–100 K \citep{Dupac_2002}. These measurements are obtained on much smaller scales than those used in this work ($<0.1$ pc). Therefore, while a $T_{\rm dust}$ of 38 K can occur in a region heated by a nearby H~{\sc ii} region, this is unlikely on 15 pc scales. On larger scales, \textit{Herschel} observations indicate a $T_{\rm dust}$ range of 15–20 K for M31 GMAs \citep{Kirk_2015}. {Thus, variations in $\alpha^{\prime}_{\rm ^{12}\rm CO}$ between clouds may reflect $T_{\rm dust}$ differences, it is also possible that some dust continuum emission is not physically associated with the GMC.}

\subsection{Virial Analysis}

Next, we performed a virial analysis of the resolved 1G dust cores in our sample {to investigate} the influence of gravity and turbulence on cloud dynamics. We include only 1G resolved sources in this analysis, {as} source radii and $^{13}\rm CO(2-1)$ linewidths are required to calculate the virial mass ($M_{\rm vir}$) following Equation \ref{eqn: Mvir}. \citet{Lada_2024} showed that, while most M31 GMCs are unbound, the strongest $^{13}\rm CO$ emission arises from bound GMCs. As discussed in Section \ref{Sec:DCP}, dust traces the densest regions within these GMCs, {therefore providing additional} insight into the physical conditions that govern star formation.

The virial theorem{, one of Larson’s relations} \citep{Larson_1981}, states that for a self-gravitating GMC, {the kinetic energy (KE) is twice the gravitational potential energy (GPE). This implies that the internal KE from turbulence is balanced by self-gravity.} Assuming a stratified cloud with an internal density gradient {of $\rho(r) \propto r^{-1}$} \citep{Solomon_1987}, {the} virial mass is then given by
\begin{equation}
\label{eqn: Mvir}
M_{\rm vir} = 1040\, \sigma^{2}R~(M_{\odot}),
\end{equation}
where $\sigma$ is the velocity dispersion in {$\rm km~s^{-1}$} and $R$ is the effective radius in pc.

For a virialised cloud{,} the virial parameter $\alpha_{\rm vir} = {2~\rm KE/GPE} = 1.0$. The virial nature of a system can be inferred from the relationship between $M_{\rm vir}$ and $M_{\rm lum}$, where a one-to-one scaling relation is expected for virialised clouds. {\citet{Larson_1981} derived} the following relation between velocity dispersion and mass: $\sigma \propto M^{0.20}$, with an rms deviation of 0.12 dex. {\citet{Lada_2024} found an almost identical relation for M31 GMCs using the same observations discussed here.} For a virialised system, $M_{\rm vir} = M_{\rm lum}^{1.0}$ is expected (e.g., \citealp{Lada_2024}).

The virial mass of each dust core is derived using $^{13}\rm CO$ linewidths for velocity dispersion{, as these are more suitable than $^{12}\rm CO$ linewidths for this purpose,} because dust and $^{13}\rm CO$ are better spatially matched in our observations (see Section \ref{Sec:maps}). $M_{\rm lum}$ is calculated using the $\alpha_{\rm ^{12}CO}$ value derived for each individual dust core and assuming $R_{\rm g/d} = 136$, multiplied by the $L_{^{12}{\rm CO}}$ measured within the dust core area. This yields the dust-based gas mass of the dust core. Because both $\alpha_{\rm CO}$ and $L_{\rm CO}$ are evaluated for each individual dust core, this calculation is mathematically equivalent to $M_{\rm lum} = 136 \times M_{\rm dust}$. Consequently, identical results are obtained when using $\alpha_{\rm ^{12}CO}$ for either $^{12}{\rm CO}$ or $^{13}{\rm CO}$.

\begin{figure}[h]
    \centering
    \includegraphics[width = 0.9\linewidth]{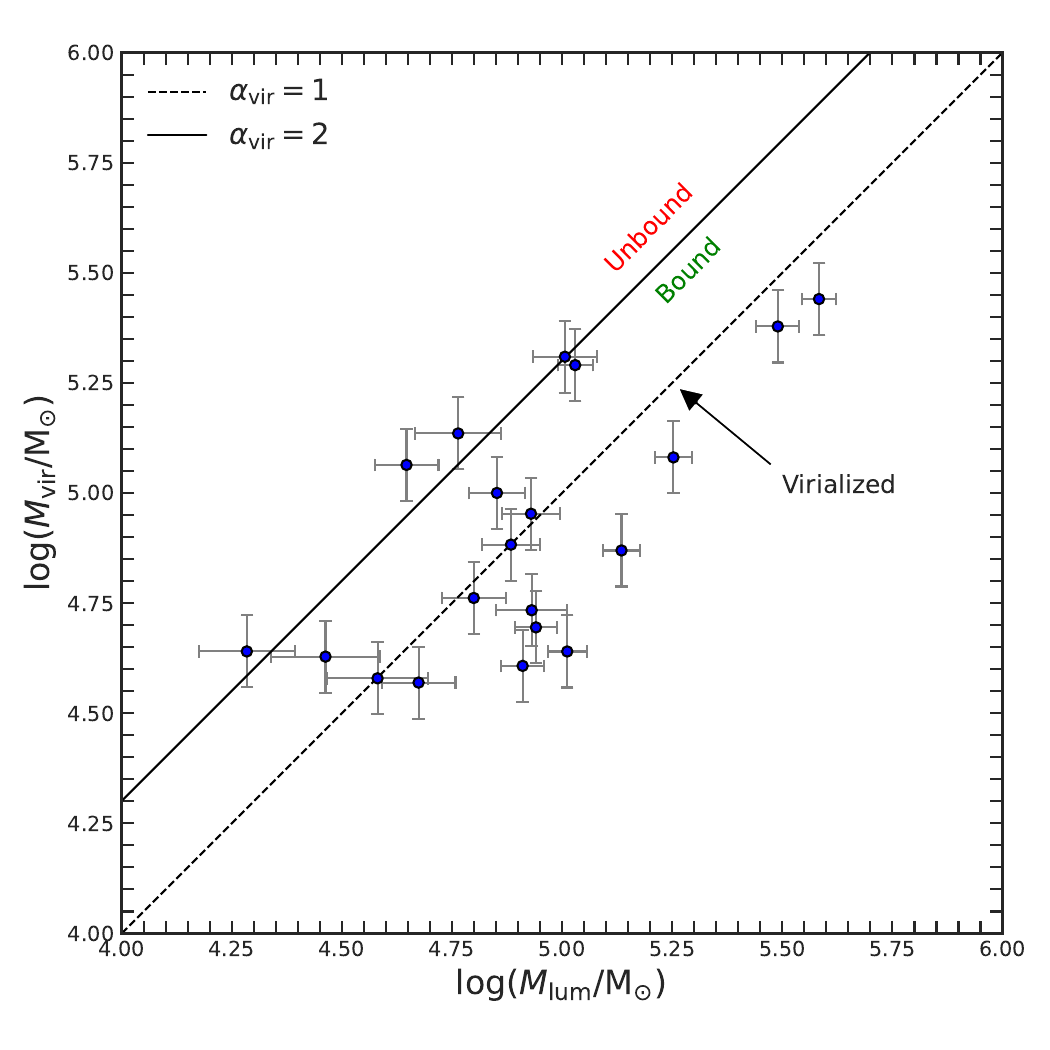}
    \caption{Comparison between the virial mass and the dust-based gas mass (both from within the 2.5$\sigma$ dust continuum contour) for individual resolved dust cores in our sample.}
    \label{fig:vir}
\end{figure}

Figure \ref{fig:vir} displays the relationship between $M_{\rm vir}$ and $M_{\rm lum}$ for our 20 resolved {1G} dust cores. The majority (80\%) of dust cores are {found to be gravitationally bound and approximately virialized}. Although four dust cores lie within the unbound area of Figure \ref{fig:vir}, {they all lie within approximately $1\sigma$ of the bound line and are therefore likely also bound}. We note that the use of $^{13}\rm CO$ linewidths for $\sigma$ provides upper limits for the $M_{\rm vir}$ of these dust cores, and ideally, $\rm C^{18}O$ linewidths would be used, {as this tracer has a lower optical depth.}

\citet{Lada_2024} {present a virial analysis} of M31 GMCs traced by $^{12}\rm CO$ and $^{13}\rm CO$ from the same dataset. Whilst only 43\% of GMCs traced by $^{12}\rm CO$ are bound, 94\% of $^{13}\rm CO$-emitting `clumps' are bound and approximately virialised {\citet{Lada_2024}}. The {results found here are consistent with those of} \cite{Lada_2024}: whilst M31 GMCs appear largely unbound, the denser regions traced by dust and $^{13}\rm CO$ are contained within bound GMCs that are also close to being in virial equilibrium. This is also similar to the results from \cite{Evans_2021} for MW GMCs, where 70–80\% of GMCs were found to be unbound, but 60\% of the $^{13}\rm CO$ regions are bound. Both \cite{Lada_2024} and \cite{Evans_2021} found that the bound fraction of GMCs increases with {cloud} mass.

\subsection{Metallicity Dependence of $\alpha^{\prime}_{\rm CO}$}
\label{Sec: met}

As introduced in Section \ref{sec:intro}, we update the result presented in \citet{Bosomworth_2025}, investigating the metallicity dependence of $\alpha^{\prime}_{\rm CO}$. This SMA sample contains 30 1G dust cores associated with the same GMCs as 30 H~{\sc ii} regions from \citet{Bosomworth_2025}. We note that the dust cores and H~{\sc ii} regions are not necessarily spatially coincident. H~{\sc ii} region metallicities were calculated from MMT/Hectospec optical spectroscopy. The primary metallicity trend found in M31 is a negative linear correlation with Galactocentric radius, as expected for spiral galaxies (e.g., \citealp{Tinsley_1980, Kewley_2019, Maiolino_2019}). $\alpha_{\rm CO}$ is predicted to be higher at low metallicities \citep{Bolatto_2013}, {and for} $\alpha^{\prime}_{\rm CO}$ if $R_{\rm g/d}$ is constant. Therefore, $\alpha_{\rm CO}$ may also be assumed to vary linearly with Galactocentric radius.

{First, both dust and CO are more abundant in higher metallicity environments. Second, the dust mass and CO luminosity are each expected to vary independently with metallicity.} Dust plays a crucial role in shielding CO from destruction by ultraviolet (UV) photons; therefore, at lower metallicities, a greater fraction of CO is destroyed by UV radiation \citep{Bolatto_2013}. {These combined effects may either exacerbate or mitigate variations in the ratio of $M_{\rm dust}$ to $L_{\rm CO}$.}

\begin{figure}[h]
    \centering
    \includegraphics[width=\linewidth]{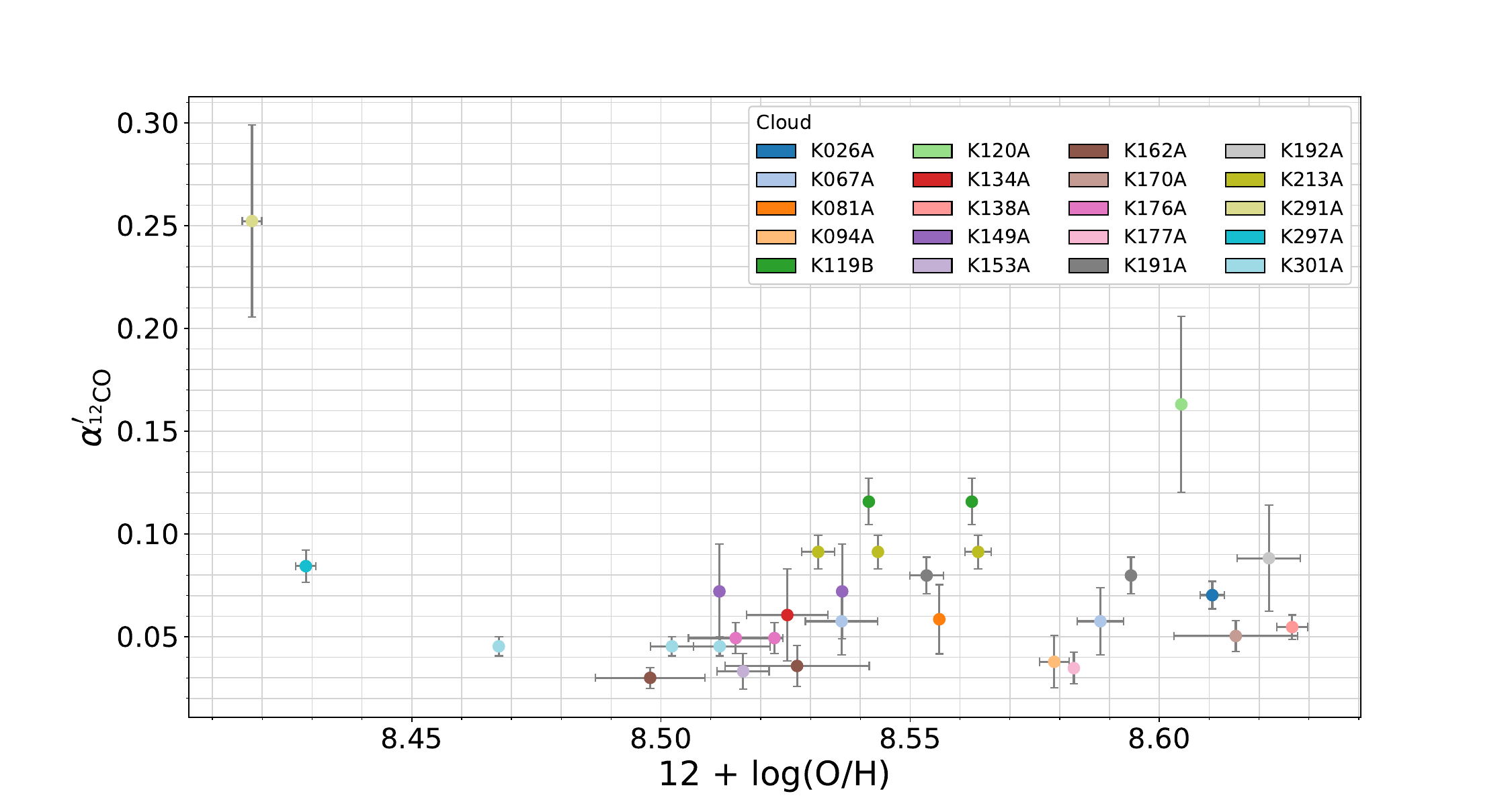}
    \caption{The variation of the CO-to-dust-mass conversion factor, $\alpha^{\prime}_{^{12}\rm CO}$, with metallicity (Oxygen abundance) for SMA dust detections in this work, which have at least one associated H~{\sc ii} region identified from optical spectroscopy \citet{Bosomworth_2025}. Some GMAs have multiple H~{\sc ii} regions, and so data points are coloured according to the associated GMA/GMC.}
    \label{fig:alphametallicity}
\end{figure}

We display the updated relationship between $\alpha^{\prime}_{\rm CO}$ and O/H in Figure \ref{fig:alphametallicity}. {Despite probing a significant metallicity range of $\sim$8.40–8.65, we find that $\alpha^{\prime}_{\rm ^{12}CO}$ remains approximately constant with a mean of $0.062 \pm 0.029~M_{\odot}\,(\rm K~km~s^{-1}~pc^{2})^{-1}$, consistent within $<2\sigma$ for most data points.} {The majority of sources in this sample have metallicities of O/H $>$ 8.5, and a significant dependence of $\alpha_{\rm CO}$ on metallicity is primarily expected at 12 + log(O/H) $<$ 8.4.} Therefore, {a larger sample at lower metallicities is required to determine whether a trend in $\alpha^{\prime}_{\rm CO}$ exists.}

Our results are consistent with a constant $\alpha^{\prime}_{\rm ^{12}CO}$ across the metallicity range 8.45 $<$ O/H $< 8.65$ in M31, and a constant $\alpha_{\rm CO}$ {if the MW value of $R_{\rm g/d} \sim 136$ is assumed}. This {contrasts with} theoretical predictions that $\alpha_{\rm CO}$ is negatively correlated with metallicity. By probing the metallicity range O/H $<$ 8.5 with a {sufficiently large} sample size, we {can better constrain} the metallicity dependence of $\alpha^{\prime}_{\rm ^{12}CO}$ in M31 and similar galaxies. Alternatively, it may be that $R_{\rm g/d}$ {varies} with metallicity.

\section{Summary \& Conclusions}
\label{sec: sum}

The recent upgrade to the SMA wideband receiver has enabled the first resolved dust continuum detections from individual GMCs in M31. Earlier studies {by} \citet{Forbrich_2020} {and} \citet{Viaene_2021} showed that {using} SMA observations of simultaneous CO(2–1) and 230 GHz continuum emission at $\sim$15~pc resolution, it is possible to derive the dust mass-to-light conversion factor ($\alpha^{\prime}_{\rm CO}$) for 32 dust cores associated with M31 GMCs (10 resolved). Analysis from the initial two observing runs of the SMA survey revealed that $\alpha^{\prime}_{\rm CO}$ for M31 GMCs is approximately constant and similar to that of the MW {when} $R_{\rm g/d} = 136$ is assumed. In this paper, we analyse the now-completed SMA survey, consisting of four observing runs targeting 80 \textit{Herschel}-identified GMAs. We increased the sample to 71 dust continuum detections (26 resolved) associated with 56 GMCs. We identified a subsample of 47 `1G' dust cores with CO line profiles well represented by a single Gaussian (20 resolved). We {calculated} $\alpha^{\prime}_{\rm CO}$ for both $\rm ^{12}CO$ and $\rm ^{13}CO$ for the 1G sample, {then performed an assessment of whether the} resolved 1G dust cores are gravitationally bound and/or virialzied. Finally, we {updated the test of the metallicity dependence of} $\alpha^{\prime}_{\rm CO}$ {using} H~{\sc ii} region metallicities presented in \citet{Bosomworth_2025}.

\begin{itemize}
    \item We re-analysed the entire dataset using a dust emission detection threshold of $2.5\sigma$ {and limited} the final sample to sources associated with GMCs defined by $\rm ^{12}CO$ emission, identifying new detections in the original two observing runs as well as in the new observations. The {defined `1G' subsample consists of sources with} single-component line profiles in both $\rm ^{12}CO(2-1)$ and $\rm ^{13}CO(2-1)$, thus eliminating confusion due to overlapping emission.
    \item We calculated the CO-to-dust mass conversion factor, $\alpha^{\prime}_{\rm CO}$, for 47 1G dust cores. The mean values derived from the sample of 20 resolved sources are {$\langle\alpha^{\prime}_{\rm ^{12}\rm CO}\rangle = 0.070 \pm 0.031~M_{\odot}\,(\rm K\,km\,s^{-1}\,pc^{2})^{-1}$} and {$\langle\alpha^{\prime}{\rm ^{13}\rm CO}\rangle = 0.37 \pm 0.20~M_{\odot}\,(\rm K\,km\,s^{-1}\,pc^{2})^{-1}$}. These values are in excellent agreement with those previously reported by \citet{Forbrich_2020} and \citet{Viaene_2021}.
    \item Assuming a MW value of $R_{\rm g/d} = 136$, the corresponding CO-to-cloud mass conversion factor is {$\langle\alpha_{^{12}\rm CO}\rangle_{\rm M31} = 9.52 \pm 4.22~M_{\odot}\,(\rm K\,km\,s^{-1}\,pc^{2})^{-1}$}. The corresponding MW value is {$\langle\alpha_{^{12}\rm CO}\rangle_{\rm MW} = 6.1~M_{\odot}\,(\rm K\,km\,s^{-1}\,pc^{2})^{-1}$}, derived from the \citet{Bolatto_2013} value for CO(1–0). {$\langle\alpha_{^{12}\rm CO}\rangle_{\rm M31}$ agrees with the MW value within $1\sigma$ uncertainties.} Our uncertainty estimate is comparable to that recommended by \citet{Bolatto_2013} for the MW, but {has a smaller relative uncertainty and is statistically derived.}
    \item We find that 80\% of the dust cores are gravitationally bound {(i.e., $\rm \vert GPE \vert < KE$)}. Moreover, all sources lie within $1\sigma$ of the gravitationally bound condition. This is consistent with the findings of \citet{Lada_2024}, who reported that M31 GMCs are largely unbound, but that dense gas regions (identified from $\rm ^{13}CO$ emission) are contained within bound GMCs. {Our results therefore suggest that the dust cores are located within these bound GMCs.}
    \item Finally, we update the \citet{Bosomworth_2025} test of $\alpha^{\prime}_{\rm CO}$ variation with metallicity (O/H). Despite the significant range of metallicities probed, we {find} no evidence of $\alpha^{\prime}_{^{12}\rm CO}$ dependence on metallicity {across the range} 8.45 $\lesssim$ O/H $\lesssim$ 8.65.
\end{itemize}

\begin{acknowledgments}

The Submillimeter Array is a joint project between the Smithsonian Astrophysical Observatory and the Academia Sinica Institute of Astronomy and Astrophysics and is funded by the Smithsonian Institution and the Academia Sinica. CB acknowledges funding from a predoctoral fellowship at Center for Astrophysics $\vert$ Harvard \& Smithsonian.
\end{acknowledgments}

\vspace{5mm}
\facilities{SMA}

\software{Astropy \citep{astropy:2013, astropy:2018, astropy:2022}, APLpy \citep{aplpy2012, aplpy2019}}

\clearpage
\appendix

\section{SMA Dust Non-Detections Source Catalog}
\label{Sec:NonDet}

\begin{figure}[!bht]
    \centering
        \includegraphics[width=0.8\linewidth]{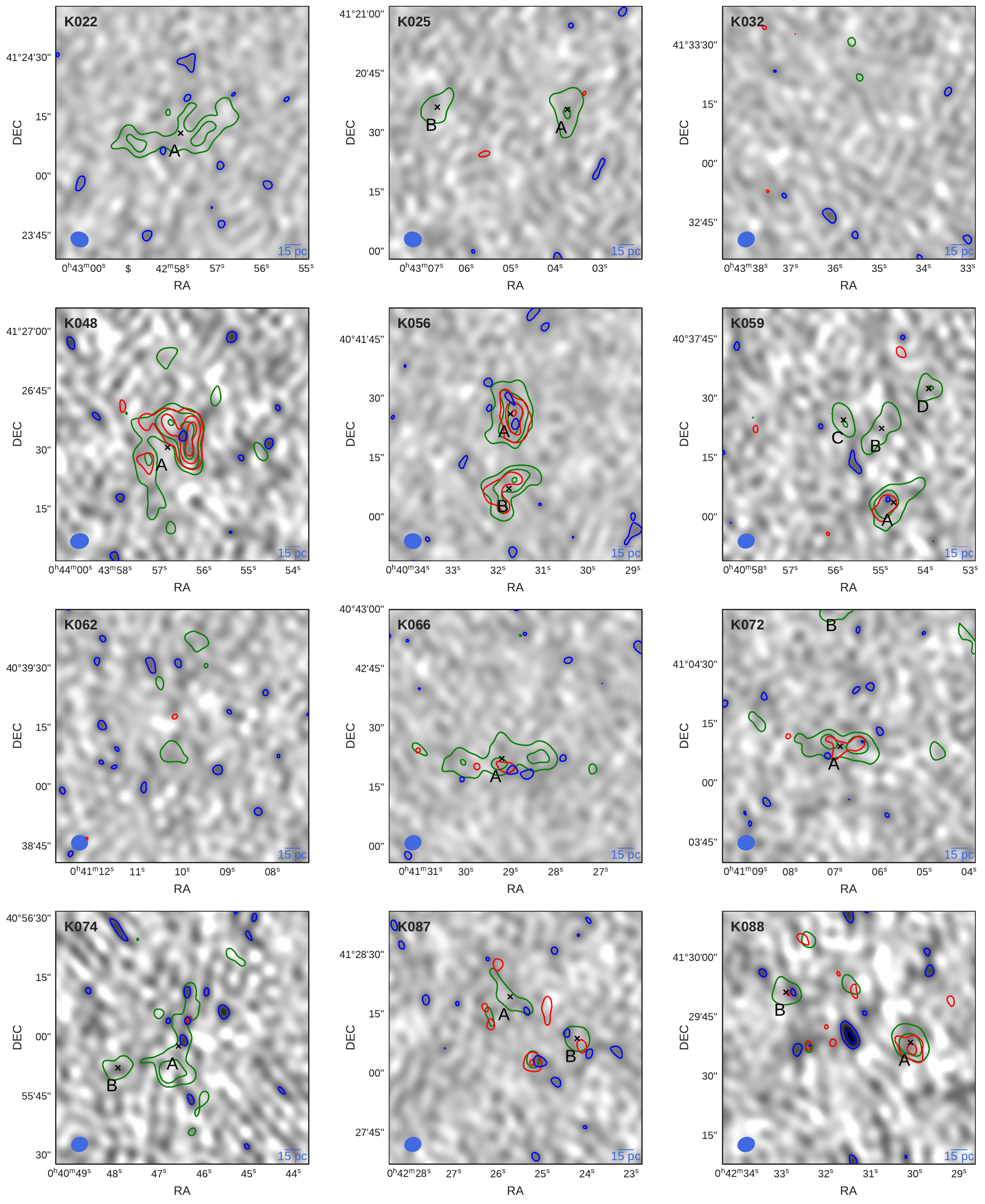}
        \caption{SMA maps of dust continuum of individual \citep{Kirk_2015} GMAs in M31, for clouds which contain no dust cores. Blue contours display dust continuum emission at 2.5, 3.5 and 4.5$\sigma$. Green contours display $\rm ^{12}CO$ at 3, 6, 12, 24 and 48$\sigma$. Red contours display $\rm ^{13}CO$ at 3, 6 and 10$\sigma$. The corresponding center of mass (by area) of individual GMCs as traced by $\rm ^{12}CO$ at 3$\sigma$ are marked by black `x's.}
        \label{fig: AtlasNon}
\end{figure}

\begin{figure}[ht]
    \centering
    \renewcommand{\thefigure}{\ref{fig: AtlasNon}}
        \includegraphics[width=0.8\linewidth]{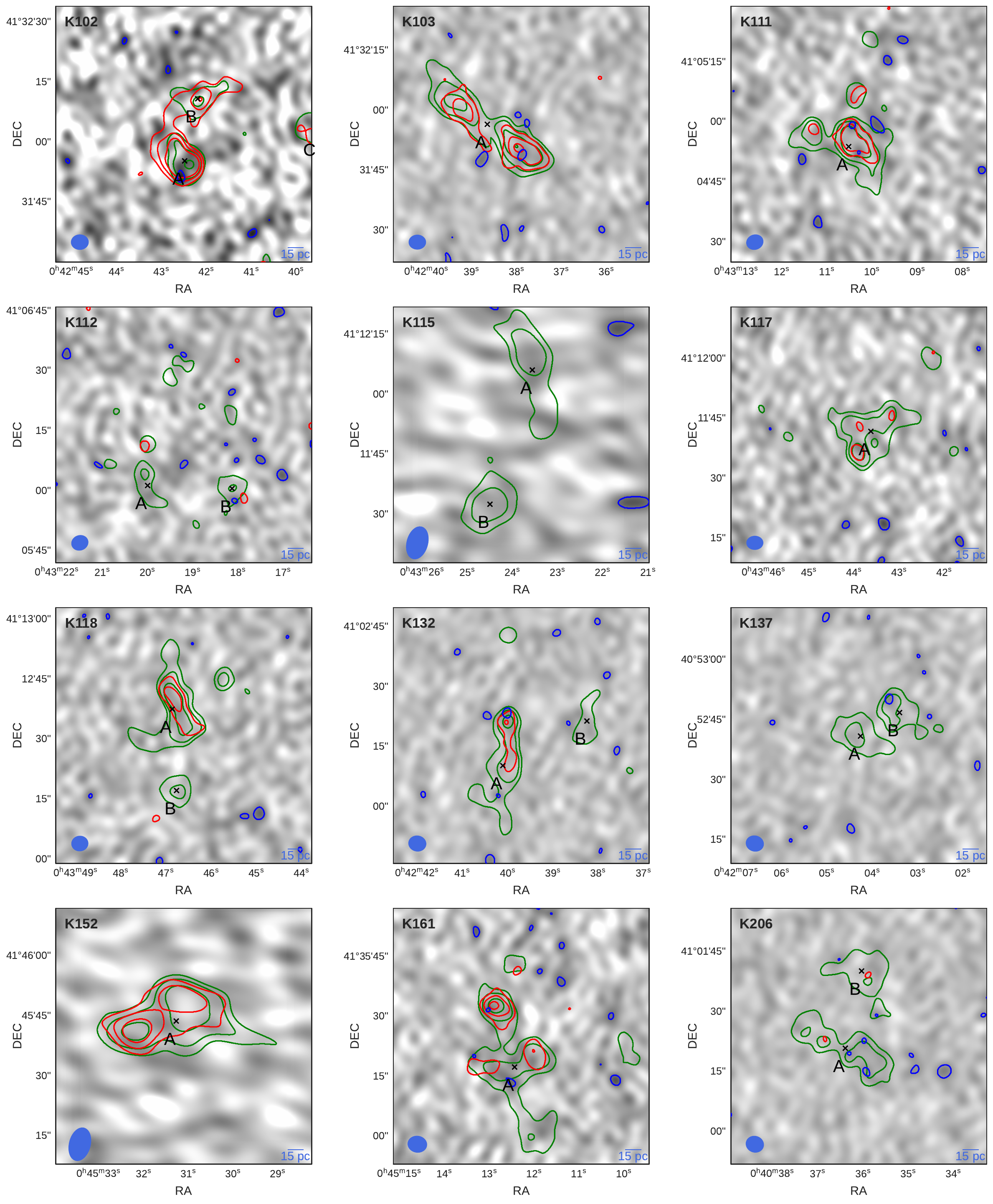}
        \caption{(continued)}
\end{figure}

\begin{figure}[ht]
    \centering
    \renewcommand{\thefigure}{\ref{fig: AtlasNon}}
        \includegraphics[width=0.55\linewidth]{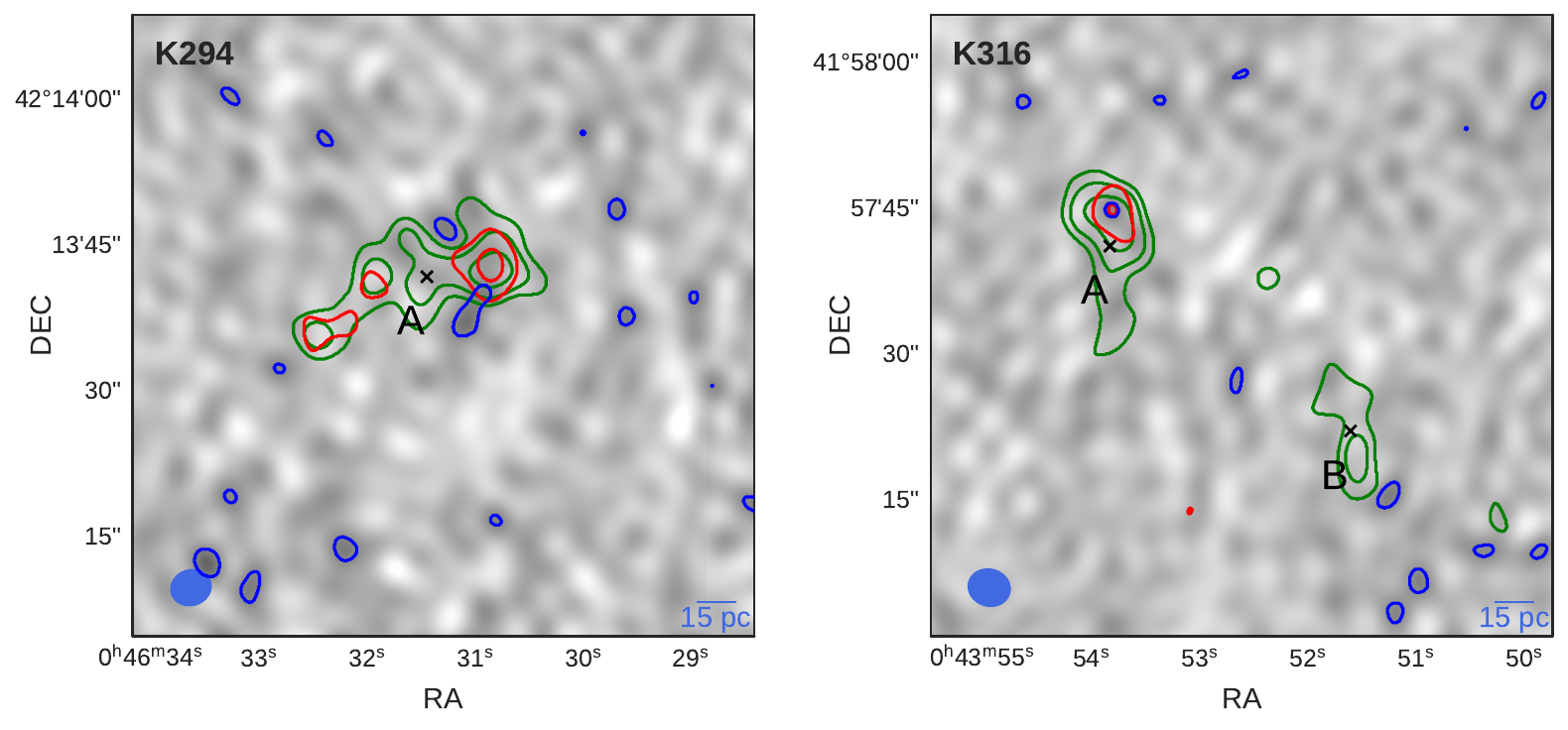}
        \caption{(continued)}
\end{figure}

\clearpage

\clearpage
\bibliographystyle{aasjournal}
\bibliography{sample631}

\end{document}